\documentclass[showpacs,preprintnumbers]{revtex4}
\usepackage{amsfonts}
\usepackage{amssymb}
\usepackage{amsmath}
\usepackage{epsfig}
\usepackage{array}
\allowdisplaybreaks
\begin{document}
\title{Study of Generalized Lema\^{\i}tre-Tolman-Bondi Spacetime in Palatini $f(R)$ Gravity}
\author{M. Z. Bhatti}
\email{mzaeem.math@pu.edu.pk}
\affiliation{Department of Mathematics, University of the Punjab, Quaid-i-Azam Campus, Lahore-54590, Pakistan.}

\author{Z. Yousaf}
\email{zeeshan.math@pu.edu.pk}
\affiliation{Department of Mathematics, University of the Punjab, Quaid-i-Azam Campus, Lahore-54590, Pakistan.}

\author{F. Hussain}
\email{furqanhussainmpf27@gmail.com}
\affiliation{Department of Mathematics, University of the Punjab, Quaid-i-Azam Campus, Lahore-54590, Pakistan.}

\begin{abstract}
This paper aims to analyze the generalization of Lema\^{\i}tre-Tolman-Bondi (LTB) spacetime for dissipative dust under the influence of Palatini $f(R)$ gravity. We explore the modified field equations, kinematical variables and mass function in this scenario. We construct Bianchi identities using conservation and differential equations for shear, expansion and curvature scalar in the background of Palatini $f(R)$ gravity. We calculated the scalar functions coming from the orthogonal decomposition of Riemann tensor in this framework. These scalar functions known as structure scalars have been explored for LTB spacetime using modified field equations. The symmetric properties of LTB spacetime have been discussed using two subcases. We found that generalized LTB spacetime has properties comparable with LTB and obtained structure scalars in both cases which have similar dependence on material profile even in Palatini gravity.
\end{abstract}
\maketitle

\section{Introduction}

General relativity (GR) is one of the most influential gravitational theory, proposed by Albert Einstein in 1915 which is based on two postulates and is a purely geometrical theory of spacetime. This theory refuted Newton's concept of gravitational force and explained the distortion of spacetime as a result of the mass-energy distribution. According to recent observations coming from cosmic microwave background radiations and type-Ia supernovae, our universe is expanding at an accelerated rate. Moreover, astronomical observations confirmed the unknown nature of dark matter present in the universe. The simplest cosmological models that describe the evolution of the universe were based on GR known as the $\Lambda$CDM model, but it does not describe the nature of dark energy. Although GR has achieved a number of milestones, but the search for dark matter and dark energy in the light of $\Lambda$CDM model required some modification. In the literature, this modification is done by modifying the gravitational part of the Einstein-Hilbert action. Modified theories such as scalar tensor theory, $f(G)$ theory, where $G$ is a Gauss-Bonnet term
\cite{nojiri2005modified}, $f(G,T)$
\cite{Yousaf2018,doi:10.1142/S021827181850044X,yousaf2019role}
(where $T$ is the trace of energy-momentum tensor), $f(R,\Box R, T)$ \cite{houndjo2017higher,doi:10.1142/S0219887818501463} (where $\Box$ is the de Alembert's operator), and
$f(R,T,R_{\gamma\delta}T^{\gamma\delta})$
\cite{haghani2013further,odintsov2013f,yousaf2017stability,Yousaf2017,yousaf2016stability},
etc. can fall in this category.

The $f(R)$ gravity has recently attracted a lot of attention due to its simplicity in describing dark energy and dark matter. The $f(R)$ theory is the simplest modification of the Einstein Hilbert action which is obtained by replacement of the Ricci scalar $R $ with its generic function. To obtain the modified field equations in this gravity, we can apply the variation in two different ways. One is the standard version in which one can vary the action integral with respect to the metric known as metric formalism, while the other is named as Palatini formalism in which the connection and metric are treated independently. The most important assumption in this formalism is that the matter action does not depend on the connection and if we ignore this assumption then we have a third formalism known as the Metric-affine formalism of $f(R)$ gravity. The most general of these theories is metric affine $f(R)$ gravity, which reduces to metric or Palatini $f(R)$ gravity under usual limits. In metric formalism, we obtain field equations of fourth-order but in the case of Palatini formalism, we have field equations of the second order which are easy to handle \cite{PhysRevD.72.083505,doi:10.1142/S0218271811018925,PhysRevD.86.044014,PhysRevD.86.104039,PhysRevD.86.127504,olmo2020stellar,olmo2020junction}.

Sotiriou \cite{sotiriou2006f} inspected a connection of $f(R)$ gravity with scalar-tensor theory using the Palatini approach. They presented a comparison between both the theories and examined the impact of their equivalence. They also have demonstrated the conditions under which matter is coupled with the independent connections. Nojiri and Odintsov \cite{nojiri2007introduction} described the problem related to accelerated expansion by taking into account some mathematical framework in $f(R)$ theory and claimed that these problems could be solved using matter coupling spacetime. Starobinsky \cite{starobinsky2007disappearing} detected a set of models for $f(R)$ gravity that produce a unique and viable cosmology. They found that these models reduced to Minkowski and de-Sitter spacetimes in the absence of matter. They also mentioned a new question concerning dark energy models on the basis of $f(R)$ gravity which has something to do with scalarons (newly discovered massive particle). Santos \emph{et al}. \cite{santos2007energy} investigated the energy conditions to discuss $f(R)$ theories of gravity and inferred the strong and null conditions with the help of the Raychaudhuri equation and other conditions using the effective energy-momentum tensor. Olmo and Alepuz \cite{olmo2011hamiltonian} analyzed the Hamiltonian composition of $f(R)$ gravity using the Palatini approach. They evaluated the constraints equation and concluded that for $\omega=-\frac{3}{2}$ the Cauchy problem is well-posed. Bamba and Geng \cite{bamba2010thermodynamics} analyzed the thermodynamics of the apparent horizon in $f(R)$ gravity using the Palatini formalism and concluded that the equilibrium framework is viable to understand the entropy of the apparent horizon. Bhatti \emph{et al}. \cite{bhatti2019stability} considered the isotropic fluid and associate them with static spherically symmetric spacetimes to examine the stability analysis of neutron star using the Palatini formalism in $f(R,T)$ gravity. They established the stable configurations under the influence of physical constraints for the spherical stars. Yousaf \cite{yousaf2020definition} modified the definition of complexity factor for the static spherically symmetric spacetime using the Palatini framework of $f(R)$ gravity. They considered the anisotropic fluid to compute the complexity factor with the help of scalar functions and studied the impact of fluid variables and $f(R)$ terms in the formation of structure with the help of the complexity factor.

Structure scalars are related to self-gravitating fluid distributions in order to understand the physical behavior of the system. Many physicists used these scalars to explain the expansion of astrophysical objects. Herrera \emph{et al}. \cite{herrera2012cylindrically} investigated that the set of stellar equations can be written in terms of scalar quantities and these quantities provide information about the basic properties of fluid distribution under consideration. Sharif and Bhatti \cite{sharif2012structure} studied the flow of dissipative fluid in the presence of a charged field. They calculated structure scalars for charged symmetry and established a relation between matter variables and curvature tensor. Sharif and Yousaf \cite{sharif2015radiating} analyzed the dynamical properties of the self-gravitating dissipative system in the background of $f(R)$ model. They found the effects of $f(R)$ terms in the composition of structure scalars and deduced that these scalars can be used to express the static solutions of $ f(R) $ field equations. Yousaf \cite{yousaf2019role} studied the impact of $f(G,T)$ terms on the scalar functions and their effects on the formation of evolution equations. They considered non-static geometry associated with anisotropic fluid and developed the $f(G,T)$ scalar functions to examine their influence in the modeling of radiating spheres. They deduced that the Penrose-Hawking singularity theorem could be studied more effectively using $f(G,T)$ structure scalars. Bhatti and Tariq \cite{bhatti2021structure} discussed the stellar equations that describe the expansion of the dissipative self-gravitating objects under the influence of charged field. They obtained a set of five structure scalars and associated them with stellar equations. They found that the scalar functions are related to fluid properties and Einstein's solutions can be written in terms of these scalars for the static case. Bhatti and his collaborators \cite{yousaf2020evolution,bhatti2020spherical,yousaf2019non} explored these scalar functions for self-gravitating objects in $f(R)$ gravity and formulated the stellar equations associated with structure scalars to study the properties of fluid distribution. They considered static solutions to reveal the consequences of scalar functions in the framework of $f(R)$ gravity.

Herrera \emph{et al}. \cite{herrera2009dynamics} discussed the Misner-Sharp technique for dissipative fluxes and associated the dynamical equations with the transport equation by including the thermodynamical effects. They compared the results with the preceding work where thermodynamics effects were ignored and discussed the consequences of these results in the field of astrophysics. Herrera \emph{et al}. \cite{herrera2014shear} studied the properties of the axially symmetric system for dissipative fluxes and concluded that the gravitational radiations are not produced in the presence of geodesic fluid. They have also mentioned that if the expansion scalar is positive then the system evolves towards Friedmann-Roberston-Walker spacetime in the absence of dissipation. Govender \emph{et al}. \cite{govender2014role} studied the impact of shear tensor on the gravitational collapse under the influence of dissipative flux and found that the existence of shear tensor has increased the core temperature using transport equation and discussed the physical properties of collapsing sphere using shear-free case. Reddy \emph{et al}. \cite{reddy2015impact} examined the influence of heat dissipation on stellar objects during the process of gravitational collapse and discussed the dynamical collapse related to Bowers and Liang's model using the perturbation scheme. They found that the temperature of the interior body is increased due to pressure anisotropy. The characteristics of different compact objects, like gravastars \cite{yousaf2020construction, yousaf2020gravastars}, wormholes \cite{sahoo2017wormholes, bambi2016wormholes} and cosmic models \cite{yadav2020existence,malik2020study,yousaf2017stellar}
are also studied in literature by various authors.

Inhomogeneous spacetimes are the exact solutions of field equations and provide information about the different generations of the universe. Lema\^{\i}tre-Tolman-Bondi (LTB) spacetime is the most common spherically symmetric inhomogeneous spacetime that plays a significant role in the study of cosmology and can be used to investigate the formation of the Cauchy and apparent horizon. Joshi and Dwivedi \cite{joshi1992structure} analyzed the inhomogeneous gravitational collapse and studied the configuration of naked singularity for adiabatic fluid using LTB models. Sussman \cite{sussman2009quasilocal} used a numerical method to extend LTB spacetime for non-zero pressure and inserted quasi-local variables to identify the dynamics of LTB metric and found initial constraints to study the effects of particle creation and gravitational collapse. Herrera \emph{et al}. \cite{herrera2011tilted} extended LTB spacetime by adding tilted congruence and examined the energy density inhomogeneities. Zibin \cite{zibin2008scalar} established the relativistic theory of linear scalar perturbations on LTB spacetime by taking into account a covariant framework. The impact of LTB geometry on higher-order derivative terms was investigated by Fernandes et al. \cite{fernandes2020high}. They studied the physical terms such as scale factor and Hubble constant for Proca structure and found that the magnetic field vanishes for this structure.

The main idea of this work is to extend the LTB metric in the background of Palatini $f(R)$ gravity
to the dissipative case. We discussed couple of aspect in this scenario. One is to explore
some basic dynamical quantities, referred as the structure scalars, which has their own physical significance.
The other is to discuss some symmetry properties of LTB metric and then generalize it under the influence of a particular modified gravity, namely Palatini $f(R)$ gravity,
along with dissipative effects,
because LTB geometry does not admit dissipation. Herrera
\cite{herrera2011physical} considered the dissipative dust fluid to
examine the impact of dissipative flux in the configuration of
energy density inhomogeneities. Sharif and Yousaf
\cite{sharif2015stability} analyzed the factors that generate the
energy density irregularities in expansion free matter distribution
by considering the dissipative and non-dissipative case in Palatini
$ f(R) $ gravity. They considered the particular case of dissipative
dust fluid and apply certain conditions to examine the contribution
of radiating parameters in the origin of energy density
irregularities.

Herrera \emph{et al}. \cite{herrera2010lemaitre}
discussed the generalization of LTB spacetime with the help of
structure scalars and symmetric properties. They considered the
dissipative dust fluid to generalize the LTB spacetime for the
dissipative case because LTB does agree with dissipative phenomena.
They have mentioned that the state of pure dust suggests that the fluid is geodesic in the absence of dissipation while this condition remains not valid when we consider dissipative flux.
Yousaf \emph{et al}. \cite{yousaf2016causes} considered the imperfect fluid to examine
the factors that create irregularities for a spherical star in $ f(R,T) $ gravity.
They studied particular cases of anisotropic, isotropic, and dust to analyze the
irregularity factors in the dissipative and non-dissipative scenarios. They considered
dissipative dust as a special case to analyze the effects of dissipation in the form of
heat flux and null radiations. Herrera \cite{herrera2015gravitational} studied the
gravitational radiation and its properties by considering dissipative fluids as well as
addressed some particular cases such as shear-free and perfect fluid to analyze
their behavior. They found that dust fluid along with dissipation is the more consistent
fluid distribution with gravitational radiation.

The Palatini version of $f(R)$ gravity have its own significance since its inception. The
theory has several fascinating properties, including the ability
to predict the existence of a long-range scalar field that explains
late-time cosmic acceleration.
It is worth noting that Olmo \cite{olmo2005post} applied the scalar-tensor representation
technique to work out the post-Newtonian limit of metric and
Palatini versions of $ f(R) $ gravity. They inferred that the $
f(R) $ Lagrangian must be linear in $ R $ in both formalisms but
observations are quite opposite to the corrections that are obtained
at low curvatures. This result demonstrated that the gravitational
alteration at very low densities is irreversible. Olmo \emph{et al}.
\cite{olmo2012stellar} used Palatini approach to examine the
formation of static spherically symmetric star, where Lagrangian is
an undefined function of $ f(R,R_{\mu\nu} R^{\mu\nu})$.
They derived the TOV equations for such theories and
indicated that under usual limits, they regain the $ f(R)$ and
general relativity. They demonstrated that exterior
vacuum solutions are comparable with Schwarzschild de-Sitter
solutions and discussed the presumed changes of the interior
solutions when matched with general relativity. Olmo and Garcia
\cite{olmo2020junction} utilized the tensor distributional technique
to compute the junction conditions for Palatini $ f(R) $ gravity.
They demonstrated that these conditions are required to construct
the models of gravitational bodies with the matching of interior and
exterior regions at hypersurface. They demonstrated their
importance by taking into account the stellar surfaces in polytropic
models and analyzed that the Palatini framework can safely model
the white dwarfs and neutron stars.

Capozziello \emph{et al}.
\cite{capozziello2012wormholes} discussed the modified theory of
gravity consisting of the
superposition of the EH Lagrangian with a Palatini term. They described the general criteria
for wormhole solutions based on null energy condition as well as
explained the red-shift function, the scalar field, and the potential that simplifies the modified Klein-Gordon equation in the first solution and matched it with vacuum solution because this solution is not asymptotically flat. They also found asymptotically flat spacetime by properly describing the metric functions and selecting the scalar field. Olmo \cite{olmo2011palatini} studied the recent work on modified theories by using the Palatini approach. He addressed various problems that have been explored within this framework after discussing the incentives that led to consider the alternatives to Einstein's theory and to treat the metric and the connection as independent variables. He discussed the cosmic speed-up
a problem in detail, as well as the solar system tests and the Cauchy problem. They also studied the importance to look beyond $ f(R)$  models to grab the phenomenological situations related to quantum gravity, dark matter and dark energy.

Amarzguioui \emph{et al}. \cite{amarzguioui2006cosmological} studied the $ f(R) $ gravity using Palatini formalism to
determine the expansion history of the universe by the choice of an arbitrary $ f(R) $
function. They explored the cosmological constraints of $ f(R) $ gravity originated from
cosmic microwave background observations and demonstrated that their results reveal that
the choice of $ \frac{1}{R} $ model is inconsistent with the data.
Allemandi \emph{et al}. \cite{allemandi2006conformal} discussed the physical significance of conformal transformations by taking the Palatini technique in gravitational theories. They analyzed that the
conformal transformations provide the physical signs which enable to differentiate
the spacetime and geodesic structures by taking into consideration the bi-metric structure within
Palatini formalism. They explained that the conformal factor is gradually changing and discussed the cosmological solutions in the Jordan and Einstein frames. Meng and Wang \cite{meng2004palatini} discussed the modified gravity with $ R^2 $
terms using the Palatini formalism. They indicated that the
quantum effects of $ R^2 $ theory are different in metric and Palatini formalisms as well
as demonstrated that $ R^2 $ term provides no information about the early time inflation
in Palatini formalism.

This paper is organized in the following manner. Section \textbf{2} includes the modified field equations and matter variables along with Bianchi identities and differential equations for Weyl and expansion scalar under the influence of Palatini $f(R)$ gravity. In section \textbf{3}, the structure scalars are found using orthogonal decomposition of Riemann tensor in the background this gravity. Section \textbf{4} includes a detailed description of LTB spacetime in which we discuss the possible solutions of the evolution equation using the graphical representation. In section \textbf{5}, the generalized LTB spacetime is obtained on the basis of structure scalars and symmetric properties. In the last section, we conclude the finding of our paper.

\section{Palatini $f(R)$ formalism}

The Einstein-Hilbert action for $ f(R)$ theory of gravity is given by
\begin{align}\label{1a}
S_{f(R)}=\frac{1}{2\kappa} \int \sqrt{-g}f(R)d^{4}x+S_{m},
\end{align}
where $ S_{M} $ denotes the action due to matter, $ g $ is the determinant of the metric tensor $ g_{\zeta\beta} $ and $ \kappa $ is the coupling constant. The scalar function $ R $ in the above action is compiled from the contraction of Ricci tensor related with connection symbols by suggesting that $ R$ is the consequence of geometrical connections.
By varying the above action integral with respect to metric $ g_{\zeta\beta} $ and connections $ \Gamma^{\mu}_{\zeta\beta} $, we acquire couple of equations as
\begin{align}\label{2a}
F(R)R_{\zeta\beta}-\frac{1}{2}g_{\zeta\beta}f(R)=&\kappa T_{\zeta\beta},
\\\label{3a}
\nabla_{\mu}(g^{\zeta\beta}\sqrt{-g}F(R))=&0,
\end{align}
where $ F(R) $ is the derivative of $ f $ with respect to $ R $ and $ T_{\zeta\beta} $ is the energy momentum tensor that does not depend on the independent connections. To make a comparison between $ T=g^{\zeta\beta} T_{\zeta\beta} $ and Ricci scalar, multiplying the above equation with conjugate metric tensor $ g^{\zeta\beta} $, we obtain the following equation
\begin{align}\label{4a}
RF(R)-2f(R)=\kappa T,
\end{align}
which describes the dependence of Ricci scalar $ R $ on the trace of energy momentum tensor.
In Palatini $ f(R) $ formalism, a single expression of the second order field equation can be obtained by calculating the connection from equation Eq.\eqref{2a} and then utilizing in Eq.\eqref{3a} as follows
\begin{align}\nonumber
R_{\zeta\beta}-\frac{1}{2}g_{\zeta\beta}R=&\frac{\kappa}{F}T_{\zeta\beta}+\frac{1}{2}g_{\zeta\beta}\left(\frac{f}{F}-R\right)+\frac{1}{F}
(\nabla_{\zeta}\nabla_{\beta}-g_{\zeta\beta}\Box F)+\frac{3}{2F^2}\\\label{5a}
&\left[\frac{1}{2}g_{\zeta\beta}(\nabla F)^{2}-\nabla_{\zeta}F \nabla_{\beta}F\right],
\end{align}
here $\nabla_{\beta}$, denotes the covariant derivative and $ \Box\equiv g^{\zeta\beta}\nabla_{\zeta} \nabla_{\beta} $ is the box function.
The Modified field equations in the form of Einstein tensor can be written as
\begin{align}\label{6a}
 G_{\zeta\beta}=\frac{8\pi}{F} (T_{\zeta\beta}+T^{(D)}_{\zeta\beta}),
\end{align}
where
\begin{align}\label{7a}
 T^{(D)}_{\zeta\beta}=&\frac{1}{\kappa} \left[\nabla_{\zeta}\nabla_{\beta} F-g_{\zeta\beta} \Box F-\frac{F}{2} g_{\zeta\beta}
R+\frac{f g_{\zeta\beta}}{2}+\frac{3}{2F}\left(\frac{1}{2}
g_{\zeta\beta}(\nabla
F)^2-\nabla_{\zeta}F\nabla_{\beta}F\right)\right],
\end{align}
is the energy-momentum tensor describing a gravitational contribution in the form of Palatini $ f(R) $ terms. The metric formalism breaks down the second-order features of GR but the Palatini approach reserves this property.
The general form of line element for the interior region of a spherically symmetric distribution of geodesic fluid can be express in diagonal form as
\begin{align}\label{8a}
ds^2=-dt^2+B^2 dr^2+C^2(d\theta^2+sin^2\theta d\phi^2),
\end{align}
where $B$ and $C$ are positive and functions of $t$ and $r$. The
metric coefficient $B$ is dimensionless while $ C $ has the same
dimension as that of $r$. It is well known that at large scales, the
cold dark matter is non-collisional and have strong influence of
rest-mass, where the dissipative and pressure terms have negligible
influence due to their kinetic nature. Therefore, it would be
significant to examine the stability of such analysis which have
influence of some kinetic terms, e.g., dissipative fluxes within the
background of a particular modified gravity namely, Palatini $f(R)$
gravity. The energy momentum tensor in this scenario is given by
\begin{align}\label{9a}
T_{\zeta\varrho}=\mu V_{\zeta} V_{\varrho}+\epsilon l_{\zeta} l_{\varrho}+q_{\zeta} V_{\varrho}+V_{\zeta} q_{\varrho},
\end{align}
here, the energy density, null four-vector and four-velocity of the
fluid are described by $ \mu,~l^{\zeta}$ and $ V_{\zeta}$
respectively. The heat flux is denoted by $ q_\zeta $
characterizing the dissipation in the form of diffusion approximation, while the $
\epsilon $ represents the radiation density defining the dissipation
in streaming-out approximation. Here, we have used the
dissipative dust fluid to discuss the dissipative phenomena for
different self-gravitating systems.
For a comoving coordinate system, the four velocity $ V^{\zeta}=\delta^{\zeta}_0$, heat flux $ q^{\zeta}=q B^{-1}\delta^{\zeta}_1$, null four vector  $ l^{\zeta}= \delta^{\zeta}_0+B^{-1}\delta^{\zeta}_1$, and unit four-vector $ \chi^{\zeta}=B^{-1} \delta^{\zeta}_1$, obey the following relations
\begin{align}\label{10a}
V^{\zeta} V_{\zeta}=&-1, \quad V^{\zeta} q_{\zeta}=0, \quad l^{\zeta} V_{\zeta}=-1, \quad l^{\zeta} l_{\zeta}=0,
 \quad \chi^{\zeta} \chi_{\zeta}=1, \quad \chi^{\zeta} V_{\zeta}=0.
\end{align}
The field equations in the background of Palatini $ f(R) $ gravity are obtained by using the
connection symbols (affinities) defined by $\Gamma^{\lambda}_{\mu\nu}=\{^{\lambda}_{\mu\nu}\}+ \frac{1}{2F} \left[( \delta^{\lambda}_{\mu} \partial_{\nu}+ \delta^{\lambda}_{\nu} \partial_{\mu})F -g_{\mu\nu} g^{\lambda\sigma} \partial_{\sigma} F  \right]$ as
 \begin{align}\label{11a}
8\pi \tilde{\mu}_{eff}=&\left(\frac{2 \dot{B}}{B}+\frac{\dot{C}}{C}\right)\frac{\dot{C}}{C}-\left(\frac{1}{B^2}\right)
\left[\frac{2C''}{C}+\left(\frac{C'}{C}\right)^2-\frac{2B' C'}{B C}-\frac{B^2}{C^2}\right]
 \\\label{12a}
-8\pi\tilde{q}_{eff} B =&-2\left(\frac{\dot{C'}}{C}-\frac{\dot{B}
C'}{B C}\right),
\\\label{13a}
8\pi B^2 \epsilon_{eff}=&-B^2\left[\frac{2\ddot{C}}{C}+\left(\frac{\dot{C}}{C}\right)^2\right]+\left(\frac{C'}{C}\right)^2-\left(\frac{B}{C}\right)^2,
\\\label{14a}
\chi_{22}=&-C^2\left(\frac{\ddot{B}}{B}+\frac{\ddot{C}}{C}+\frac{\dot{B}\dot{C}}{B
C}\right)+\left(\frac{C}{B}\right)^2\left(\frac{C''}{C}-\frac{B'
C'}{B C}\right),
\end{align}
where $ \tilde{\mu}_{eff}, ~\tilde{q}_{eff}$ and $ \epsilon_{eff} $
contain the $ f(R) $ corrections along  with usual terms and their
values are given by
\begin{align}\label{15a}
\tilde{\mu}_{eff}=\frac{1}{F} (\tilde{\mu}+\chi_{00}), \quad
\tilde{q}_{eff}=\frac{1}{F}\left(\tilde{q}+\frac{\chi_{01}}{B}\right),
\quad
\epsilon_{eff}=\frac{1}{F}\left(\epsilon+\frac{\chi_{11}}{B^2}\right),
 \end{align}
here $\tilde{\mu}=\mu+\epsilon$ and $\tilde{q}=q+\epsilon$, while $
\chi_{00},~\chi_{01},~\chi_{11} $ and $ \chi_{22} $ represent the
dark source terms whose values are given in the Appendix A. Here dot
stands for $t$ differentiation and prime stands for differentiation
w.r.t $r$ coordinate.

The expansion scalar $ \Theta $, four acceleration $ a_{\zeta} $ and shear tensor $ \sigma_{\zeta\varrho} $ can be defined as
\begin{align}\label{16a}
\Theta= V^{\zeta};_{\zeta} \quad a_{\zeta}=V_{\zeta;\varrho} V^{\varrho} \quad  \sigma_{\zeta\varrho}=V_{({\zeta;\varrho})}-\frac{1}{3} \Theta h_{\zeta\varrho}+a_{(\zeta}V_{\varrho})
\end{align}
where $ (h_{\zeta\varrho}=g_{\zeta\varrho}+V_{\zeta} V_{\varrho}) $. The expansion scalar $ \Theta $ and the non vanishing elements of four acceleration $ a_{\zeta} $  and shear tensor $ \sigma_{\zeta\varrho} $ in $ f(R) $ gravity are given by
\begin{align}\label{17a}
\Theta_{eff}=&\frac{\dot{B}}{B}+2\frac{\dot{C}}{C}+2\frac{\dot{F}}{F},\quad a_{0}=\frac{\dot{F}}{2F}, \quad a_{1}=\frac{F'}{2F},
\\\label{18a}
 \sigma_{01}=&\frac{F'}{4F}, \quad \sigma_{11}=\frac{2}{3} B^2 \sigma-\frac{\dot{F}B^2}{6F}, \quad \sigma_{22}=\frac{\sigma_{33}}{sin^2 \theta}=
 -\frac{1}{3} C^2 \sigma-\frac{\dot{F}C^2}{6F},
\end{align}
The presence of four acceleration is totally dependent on Palatini $ f(R)$ formalism as observed from Eq.\eqref{17a}. The shear scalar is defined as $ \sigma=\frac{\dot{B}}{B}-\frac{\dot{C}}{C}$, and its value in Palatini $ f(R) $ gravity turns out to be
\begin{align}\label{19a}
\sigma_{eff}=\left(\sigma^2+\frac{\dot{F}^2}{8F^2}-\frac{3F'^2}{16F^2 B^2}\right)^{\frac{1}{2}}
\end{align}
Misner and Sharp \cite{misner1964relativistic} presented a formula for mass function $ m(t,r) $ to measure the amount of matter inside a self-gravitating system which for our line element takes the form as
\begin{align}\label{20a}
m=\frac{(C^3)}{2} R_{23}^{~~23}=\frac{C}{2}\left[\dot{C}^2-\left(\frac{C'}{B}\right)^2+1\right].
\end{align}
The above equation can be written as
\begin{align}\label{21a}
E=\frac{R'}{B}=\left[1+U^2-\frac{2m(t,r)}{R}\right]^{\frac{1}{2}},
\end{align}
where $ U $ is collapsing fluid velocity and can be defined as the variation of areal radius $ C $ with its proper time. In case of collapsing spheres it is taken to be negative.
Using the above expression, we can write Eq.\eqref{12a} as
\begin{align}\label{22a}
4\pi \tilde{q}_{eff}=E \left[\frac{1}{3C'}(\Theta-\sigma)'-\frac{\sigma}{C}\right].
\end{align}
The radial and temporal variation of the mass function leads to the following couple of equations as
\begin{align}\label{23a}
\dot{m}=&-4\pi(\epsilon_{eff} U + \tilde{q}_{eff} E) C^2,
\\\label{24a}
m'=&4\pi\left(\tilde{\mu}_{eff}+\tilde{q}_{eff}\frac{U}{E}\right)C^2 C',
\end{align}
where $ \tilde{q}=q+\epsilon $ and $ \tilde{\mu}=\mu+\epsilon $. Integrating partially the Eq.\eqref{24a}, we get
\begin{align}\label{25a}
\frac{3m}{C^3}=4\pi \tilde{\mu}_{eff}-\frac{4\pi}{C^3} \int_0^r  C^3 \left(\tilde{\mu}_{eff}'-3\tilde{q}_{eff}\frac{C' U}{C E}\right)dr.
\end{align}
This equation provide a relationship between mass function and matter variables such as heat flux and energy density in the presence of Palatini $ f(R) $ terms.

\subsection{Matching conditions}

To to obtain matching conditions for exterior and interior
geometry, we consider the exterior metric of the following form
\cite{olmo2005post}
\begin{align}\label{26a}
ds^{2}=-\left(1-\frac{2M}{r}-\frac{\tilde{R}r^2}{12}\right)d\nu^2-2d\nu
dr+r^2 d\theta^2 + r^2 sin^{2}\theta d\phi^2,
\end{align}
where total mass of the fluid is represented by $ M $ and $ \nu $
denotes the retarded time. To develop the smooth matching on the
boundary surface $ r= r_{\Sigma} $ for interior and exterior
geometry, we use Darmois as well as Senovilla \cite{PhysRevD.88.064015} junction conditions which in the presence
of Palatini $f(R)$ gravity give \cite{sharif2014dynamical}
\begin{align}\label{27a}
 M+\frac{r^3 R}{24}& \overset \Sigma= m(t,r),
\\\label{28a}
\frac{R r}{12}-\frac{R}{2 \kappa} \left(f_{R}-\frac{f}{R}\right)&
\overset \Sigma=q_{eff}.
\end{align}
Further, we have
$R|^+_-=0,\quad f_{,RR} \partial_\nu R|^+_-=0,\quad f_{,RR}\neq0$, which is required to be
obeyed over the boundary due to modified gravity which assures the continuity of Ricci
curvature invariant over the hypersurface even for matter thin shells.
Moreover, we can obtain over the hypersurface
\begin{align}\label{29a}
\epsilon_{eff}\overset{\Sigma}=\frac{L}{4\pi\rho^2},
\end{align}
where $ L $ represents the total luminosity of the sphere calculated on boundary surface $ \Sigma $, and has the form
\begin{align}\label{30a}
L\overset{\Sigma}=L_{\infty} \left(1-+2\frac{d\rho}{d\nu}-\frac{2m}{\rho}\right)^{-1},
\end{align}
where $ L_{\infty} $ is the total luminosity determined by a stationary observer at infinity as
\begin{align}\label{31a}
L_{\infty}=\frac{dM}{d\nu}\overset{\Sigma}=-\left[\frac{dm}{dt}\left(\frac{d\nu}{dt}\right)^{-1}\right],
\end{align}
The red-shift $ z_{\Sigma} $ on the boundary surface is given by
\begin{align}\label{32a}
\frac{d\nu}{dt}\overset{\Sigma}=1+z,
\end{align}
here
\begin{align}\label{33a}
\frac{d\nu}{dt}\overset{\Sigma}=\left(\frac{C'}{B}+\dot{C}\right)^{-1}.
\end{align}
Therefore, the time formation of the black hole is
\begin{align}\label{34a}
\left(\frac{C'}{B}+\dot{C}\right)\overset{\Sigma}=E+U \overset{\Sigma}=0.
\end{align}
Also, using Eqs.\eqref{30a} and \eqref{33a}, we find
\begin{align}\label{35a}
L\overset{\Sigma}=\frac{L_{\infty}}{(E+U)^{2}},
\end{align}
while from Eqs.\eqref{21a}, and \eqref{33a}, it follows
\begin{align}\label{36a}
\frac{d\rho}{d\nu}\overset{\Sigma}=U(U+E).
\end{align}
At the end, it is worthy to mention that the total luminosity $ L_{\Sigma} $ disappear in the case of heat flux $ (\epsilon_{eff}=0, q_{eff}\neq0) $.

\subsection{Relationship between the Weyl scalar and matter variables}

The Weyl tensor $ C^{\rho}_{\zeta\varrho\mu}$ can be expressed in terms of  Riemann tensor $ R^{\rho}_{\zeta\varrho\mu}, $ the Ricci tensor $ R_{\zeta\varrho} $ and the curvature scalar $ R $ as
\begin{align}\label{37a}
C^{\rho}_{\zeta\varrho\mu}=&R^{\rho}_{\zeta\varrho\mu}-\frac{1}{2}R^{\rho}_{\varrho}g_{\zeta\mu}+\frac{1}{2}R_{\zeta\varrho}\delta^{\rho}_{\mu}-\frac{1}{2}R_{\zeta\mu} \delta^{\rho}_{\varrho}+\frac{1}{2}R^{\rho}_{\mu} g_{\zeta\varrho}+\frac{1}{6}R(\delta^{\rho}_{\varrho} g_{\zeta\mu}-g_{\zeta\varrho} \delta^{\rho}_{\mu}),
\end{align}
which can be decomposed into its magnetic and electric parts. For spherically symmetric distribution of matter, the magnetic part of the Weyl tensor vanishes and it can only be expressed in terms of its electric part stated as
\begin{align}\label{38a}
E_{\zeta\varrho}=C_{\zeta\mu\varrho\nu} V^{\mu} V^{\nu},
\end{align}
The electric part of the Weyl tensor in terms of unit four-vector and projection tensor can be written as
\begin{align}\label{39a}
E_{\zeta\varrho}=\varepsilon (\chi_{\zeta} \chi_{\varrho}-\frac{1}{3} h_{\zeta\varrho}),
\end{align}
where
\begin{align}\label{40a}
\varepsilon=&\frac{1}{2}\left[\frac{\ddot{C}}{C}-\frac{\ddot{B}}{B}-\left(\frac{\dot{C}}{C}-\frac{\dot{B}}{B}\right)\frac{\dot{C}}{C}\right]
+\frac{1}{2B^2}\left[-\frac{C''}{C}+\left(\frac{B'}{B}+\frac{C'}{C}\right)\frac{C'}{C}\right]-\frac{1}{2C^2}.
\end{align}
is the scalar enclosing the effects of spacetime curvature. Now,
using Eqs.\eqref{11a}, \eqref{13a}, \eqref{14a}, \eqref{20a} and
\eqref{40a}, we have
\begin{align}\label{41a}
4\pi(\tilde{\mu}_{eff}-\epsilon_{eff})-\varepsilon=\frac{3m}{C^3}-\frac{\chi_{22}}{C^2}.
\end{align}
The above equation shows the dependence of the Weyl scalar on the fluid variables and mass function in the presence of dark source terms.
The differential equations of expansion and shear scalar describing the motion and expansion of particles respectively, can be read as
\begin{align}\label{42a}
\dot{\Theta}_{eff}+\frac{1}{3} \Theta_{eff}^2+\frac{2}{3}
\sigma_{eff}^2=&-4\pi(\tilde{\mu}_{eff}+\epsilon_{eff})-\frac{\chi_{22}}{C^2}+D_{1},
\\\label{43a}
\dot{\sigma}_{eff}+\frac{1}{3} \sigma_{eff}^2+\frac{2}{3} \Theta_{eff} \sigma_{eff}=&T_{1}(4\pi \epsilon_{eff}-\varepsilon+T_{2})+D_{2},
\end{align}
here, $ T_{1}=2\left(\frac{\dot{B}}{B}-\frac{\dot{C}}{C}\right) $ and  $T_{2}=2\frac{\dot{B}\dot{C}}{BC}-\frac{\dot{B}^2}{3B^2}-\frac{\dot{C}^2}{3C^2}$,
where $ D_{1} $ and $ D_{2} $ appear due to Palatini formalism of $ f(R) $ gravity, whose values are in appendix C.
The following equations derive from the conservation of energy and momentum using matter fields in the framework of the Palatini $ f(R) $ gravity as
\begin{align}\label{44a}
&\dot{\tilde{\mu}}_{eff}+(\tilde{\mu}_{eff}+\epsilon_{eff})\frac{\dot{B}}{B}+\frac{2\dot{C}}{C} \tilde{\mu}_{eff}+\frac{\tilde{q}'_{eff}}{B}+2\frac{C'}{C B} \tilde{q}_{eff}+\frac{5\dot{F}}{2F} \tilde{\mu}_{eff}+\frac{3F'}{F B} \tilde{q}_{eff}+\frac{\dot{F}}{2F} \epsilon_{eff}=0,
\\\label{45a}
&\dot{\tilde{q}}_{eff}+2 \left(\frac{\dot{B}}{B}+\frac{\dot{C}}{C}\right) \tilde{q}_{eff}+\frac{\epsilon'_{eff}}{B}+\frac{2C'}{C B} \epsilon_{eff}+\frac{3\dot{F}}{F} \tilde{q}_{eff}+\frac{5F'}{2F B} \epsilon_{eff}+\frac{F'}{2F B}\tilde{\mu}_{eff}=0.
\end{align}
The Bianchi identities can be used to obtain the differential equation for the Weyl tensor as
\begin{align}\label{46a}
[\varepsilon-4\pi(\tilde{\mu}_{eff}-\epsilon_{eff})
]'+(\varepsilon+4\pi \epsilon_{eff} )\frac{3C'}{C}=&-12\pi B
\frac{\dot{C}}{C} \tilde{q}_{eff}+
\frac{\chi'_{22}}{2C^2}+\frac{\chi_{22} C'}{2C^3}.
\end{align}
It describes a relation between the Weyl tensor and matter variables in the presence of extra curvature terms.

\section{Palatini Structure Scalars}

Bel \cite{bel1961inductions} did the pioneer work to suggest the decomposition of the Riemann tensor $ R_{\zeta\gamma\varrho\delta} $. It is worth significant to mention here that Herrera \emph{et al}. \cite{herrera2009structure} followed the Bel
\cite{bel1961inductions} procedure to split the Riemann tensor
orthogonally to associate the properties of the fluid
distribution with structure scalars in the context of general relativity. They
introduced the magnetic $ Z_{\alpha\beta}$, electric $
Y_{\alpha\beta}$, and dual part $ X_{\alpha\beta} $ of the Riemann
tensor, which is obtained from the orthogonal splitting of the
Riemann tensor. These tensors are further split into their trace and
trace free parts to obtain the five structure scalars (for the case of spherical spacetime) as well as
described the physical meaning of each structure scalar. The
scalars $ Y_{T} $ and $ X_{T} $ gives information about pressure
anisotropy and the energy density of the fluid, respectively. The
scalar $ Y_{T F} $ comprehends the pressure anisotropy with addition of
the Weyl scalar, while the inhomogeneities produced in energy density is managed by
the scalar $ X_{T F} $ in the scarcity of dissipation.
Later on, different
authors \cite{herrera2012cylindrically, sharif2012structure} have used the structure
scalars for the study of cylindrically symmetric fluids with and without considering the
electromagnetic effects as well as discussed their physical significance.

Bhatti \emph{et al}. \cite{bhatti2021role, bhatti2021analysis}
calculated the scalar functions and discussed their role on the evolution of compact objects using Palatini $ f(R) $ gravity in the presence
of the electromagnetic field. They evaluated the kinematical variables to examine the physical features of the fluid as well as Bianchi identities and the Raychaudhuri equation. They break the Riemann tensor orthogonally to get scalar functions and associated them with physical properties of the fluid like energy density and anisotropic stresses as well as explained that these Palatini scalar functions enable us to indicate the formation of the singularity in celestial objects. Yousaf \cite{yousaf2020definition} computed the Palatini structure scalars to modify the definition of complexity in $ f(R) $ gravity as well as investigated the field equations and TOV equation by taking anisotropic matter distribution. We observed that complexity is defined through Palatini structure scalar $ Y_{T F} $ in $ f(R) $ gravity with the exception that the non-attractive Palatini $ f(R) $ terms in $ Y_{T F} $ slow down the transition and supplying impendence to the system that prevent its homogeneous condition. Here, we define the tensors $ Y_{\zeta\varrho} $  and $ X_{\zeta\varrho} $ which are the components of the orthogonal decomposition in order to obtain a formalism in Palatini $ f(R) $ gravity for structure scalars.
\begin{align}\label{47a}
Y_{\zeta\varrho}=&R_{\zeta\gamma\varrho\delta} V^{\gamma} V^{\delta}, \quad  X_{\zeta\varrho}=~^{*}R^{\ast}_{\zeta\gamma\varrho\delta} V^{\gamma} V^{\delta}=\frac{1}{2} \eta_{\zeta\gamma}^{~~\epsilon\rho} R^{\ast}_{\epsilon\rho\varrho\delta} V^{\gamma} V^{\delta},
\end{align}
where $ R^{\ast}_{\zeta\varrho\gamma\delta}=\frac{1}{2} \eta_{\epsilon\gamma\rho\delta} R_{\zeta\varrho}^{~~\epsilon\rho} $ and $ ~^{*}R^{\ast}_{\zeta\gamma\varrho\delta}=\frac{1}{2} \eta_{\zeta\gamma}^{~~\epsilon\rho} R^{\ast}_{\epsilon\rho\varrho\delta} $ represent the right and double dual of the Riemann tensor and $ \eta_{\epsilon\rho\gamma\delta} $ represents the Levi-Civita tensor.
Using the natural way of the orthogonal decomposition for Riemann tensor, $Y_{\zeta\varrho} $  and $ X_{\zeta\varrho} $  which contain the physical variables and some dark source terms of modified gravity can be expressed as given in Appendix D.
These tensors can further be divided into some scalar quantities, which are known as structure scalars. Consequently, we can find $ Y_{T}, Y_{T F}, X_{T} $ and $ X_{T F} $ as
\begin{align}\label{48a}
Y_{T}=&4\pi(\tilde{\mu}_{eff}+\epsilon_{eff})+M^{(D)}_{1},
\\\label{49a}
Y_{T F}=&\varepsilon-4\pi \epsilon_{eff}+\frac{4\pi}{F} \chi_{11}+M^{(D)}_{2},
\\\label{50a}
X_{T}=&8\pi \tilde{\mu}_{eff}-\frac{8\pi}{F} \chi_{00}+M^{(D)}_{3},
\\\label{51a}
X_{T F}=&-\varepsilon-4\pi \epsilon_{eff} +\frac{4\pi}{F} \chi_{11}+M^{(D)}_{4},
\end{align}
where
\begin{align}\label{52a}
 M^{(D)}_{2}=\frac{M^{(D)}_{\zeta\varrho}}{\chi_{\zeta} \chi_{\varrho}-\frac{1}{3} h_{\zeta\varrho}} \quad  M^{(D)}_{4}=\frac{N^{(D)}_{\zeta\varrho}}{\chi_{\zeta} \chi_{\varrho}-\frac{1}{3}h_{\zeta\varrho}}.
\end{align}
By making use of Eqs.\eqref{25a}, \eqref{41a} and \eqref{49a}, we
obtain
\begin{align}\label{53a}
Y_{T F}=-8\pi \epsilon_{eff}+\frac{4\pi}{C^3} \int C^3\left(\tilde{\mu}'_{eff}-3\tilde{q}_{eff}\frac{C' U}{C E}\right)dr+\frac{\chi_{22}}{C^2}.
\end{align}
The values of $ M^{(D)}_{1},~M^{(D)}_{3},~M^{(D)}_{\zeta\varrho}$
and $ N^{(D)}_{\zeta\varrho} $ are given by in the Appendix B. The
scalar $ Y_{TF} $ can be expressed in terms of the null radiation,
energy density inhomogeneity, and the dissipative variables along
with some modified correction terms \cite{herrera2009structure}.
Furthermore, it has been demonstrated that $ Y_{TF} $ manage the
stability of the shear-free condition in the geodesic case. The
values of structure scalars obtained in Eqs.\eqref{48a}-\eqref{51a}
are formulated in terms of geometrical variables as
\begin{align}\label{54a}
Y_{T}=&-2\frac{\ddot{C}}{C}-\frac{\ddot{B}}{B}-\frac{\chi_{22}}{C^2}+M^{(D)}_{1},
\\\label{55a}
Y_{T F}=&-\frac{\ddot{B}}{B}+\frac{\ddot{C}}{C}-\frac{\chi_{22}}{C^2}+\frac{4\pi}{F} \chi_{11}+M^{(D)}_{2},
\\\label{56a}
X_{T}=&\left(2\frac{\dot{B}}{B}+\frac{\dot{C}}{C}\right)\frac{\dot{C}}{C}-\frac{1}{B^2}\left[2\frac{C''}{C}+\frac{C'^2}{C^2}-2\frac{B' C'}{B C}\right]+\frac{1}{C^2}-\frac{8\pi}{F} \chi_{00}+M^{(D)}_{3},
\\\label{57a}
X_{T F}=&\left(\frac{\dot{C}}{C}-\frac{\dot{B}}{B}\right)\frac{\dot{C}}{C}+\frac{1}{B^2}\left[\frac{C''}{C}-\left(\frac{B'}{B}+\frac{C'}{C}\right)
\frac{C'}{C}\right]+\frac{1}{C^2}-\frac{\chi_{22}}{2C^2}+\frac{4\pi}{F} \chi_{11}+M^{(D)}_{4}.
\end{align}
The GR structure scalars \cite{herrera2010lemaitre} can be retrieved
directly from the above expressions in the absence of $f(R)$
curvature terms. In the modeling of compact objects in $f(R)$
gravity, these scalar functions are closely linked. In order to
investigate the effects of $ \epsilon R^2 $ in this context, we use
the Raychaudhuri equation as a useful tool. In terms of these
scalars functions, Equations \eqref{42a}, \eqref{43a} and
\eqref{46a} become
\begin{align}\label{58a}
\dot{\Theta}_{eff}+\frac{1}{3} \Theta^2_{eff}+\frac{2}{3} \sigma^2_{eff}=&-Y_{T}+M^{(D)}_{1}-\frac{\chi_{3}}{C^{2}}+D_{1},
\end{align}
This gives a well-discussed expansion evolution of the system with the assistance scalar function $ Y_{T} $ in the presence of extra curvature terms of $f(R)$ gravity. The equation that explains the shearing motion in term of structure scalar is given by
\begin{align}\label{59a}
\dot{\sigma}_{eff}+\frac{1}{3} \sigma^2_{eff}+\frac{2}{3} \Theta_{eff} \sigma_{eff}=&T_{1}(-Y_{T F}+M^{(D)}_{2}+\frac{4\pi}{F}\chi_{11}+T_{2})+D_{2},
\end{align}
As a result, above equation shows that shearing motion is affected due to Palatini $ f(R) $ dark source terms with spherical spacetime. Furthermore, the differential equation describing the Weyl tensor's relationship with matter variables in terms of structure scalars and extra curvature terms is obtained as
\begin{align}\nonumber
(X_{T F}+4\pi \tilde{\mu}_{eff})'=&-3\frac{C'}{C} X_{T F} +4\pi
\tilde{q}_{eff}(\Theta-\sigma)
B-\frac{\chi'_{22}}{2C^2}+\frac{\chi_{22} C'}{2C^3}+
\left(\frac{4\pi}{F} \chi_{11}+M_{4}^{(D)}\right)'\\\label{60a}
&+\left(\frac{4\pi}{F} \chi_{11}+M_{4}^{(D)}\right) \frac{3C'}{C}.
\end{align}
In case of non-dissipative dust cloud, Eq.\eqref{60a} reduces to
\begin{align}\label{61a}
(X_{T F}+4\pi \tilde{\mu}_{eff})'=&-3\frac{C'}{C} X_{T F}-\frac{\chi'_{22}}{2C^2}+\frac{\chi_{22} C'}{2C^3}+\left(\frac{4\pi}{F} \chi_{11}+M_{4}^{(D)}\right)'+\left(\frac{4\pi}{F} \chi_{11}+M_{4}^{(D)}\right) \frac{3C'}{C}.
\end{align}
When $ \tilde{\mu}'_{eff}=0 $ in the above equation, then it yields
\begin{align}\label{62a}
(X_{T F})' =&-3\frac{C'}{C} X_{T F}-\frac{\chi'_{22}}{2C^2}+\frac{\chi_{22} C'}{2C^3}+\left(\frac{4\pi}{F} \chi_{11}+M_{4}^{(D)}\right)'
+\left(\frac{4\pi}{F} \chi_{11}+M_{4}^{(D)}\right) \frac{3C'}{C},
\end{align}
whose integration leads to
\begin{align}\label{63a}
X_{T F} =&\frac{f(t)}{C^3}-\int \frac{1}{X_{T F}}\left[\frac{\chi'_{22}}{2C^2}+\frac{\chi_{22} C'}{2C^3}+\left(\frac{4\pi}{F} \chi_{11}+M_{4}^{(D)}\right)'+\left(\frac{4\pi}{F} \chi_{11}+M_{4}^{(D)}\right) \frac{3C'}{C}\right]dr.
\end{align}
It should be clear from Eq.\eqref{63a} that the scalar $ X_{T F} $
controls the energy density inhomogeneity in case of non-dissipation
and in the absence of $ \epsilon_{eff} $ terms. This consequence is
also held for an isotropic fluid established in
\cite{herrera2009structure}.

\section{Lema\^{\i}tre-Tolman-Bondi metric}

In this section, we discuss the properties of LTB geometry for non-dissipative dust fluid and obtain different solutions of the evolution equation in the case of LTB. We suppose a geodesic, non-dissipative fluid $ ( q_{eff}=\epsilon_{eff}=0 ) $ for the purpose of obtaining a general structure of LTB model. In this scenario, integration of Eq.\eqref{12a} yields
\begin{align}\label{64a}
B(t,r)=\frac{1}{1+k(r)}
\exp\left[\int\frac{\frac{1}{2F}\left(\dot{F'}-\frac{5\dot{F}F'}{2F}\right)-
\frac{\dot{C'}}{C}}{\left(\frac{F'}{2F}-\frac{C'}{C}\right)}dt\right],
\end{align}
where  $ \kappa $ is an integration constant, while in case of metric $ f(R) $ gravity the term $
\frac{5\dot{F}F'}{2F} $ is absent because it appears due to
connection symbols. In case of GR, $ F\longrightarrow 1$,
 $ \dot{F},~F'\longrightarrow 0$  and above expression reduces to
\begin{align}\nonumber
B(t,r)=\frac{C'(t,r)}{(1+k(r))^{\frac{1}{2}}}.
\end{align}
The evolution equation in case of Palatini $ f(R)$ formalism is
obtained by comparing Eq.\eqref{64a} with Misner-Sharp mass given in
Eq.(20). In order to obtain the solution of evolution equation, we
use $ R+\alpha R^2 $ (Starobinski) model and junction conditions.
The equation is numerically solved and its solution is described
through graphical representation which describe the effects of $
f(R) $ gravity on different models of the universe and expressed in
Figures $1,~2 $ and $3$.
\begin{figure}
\epsfig{file=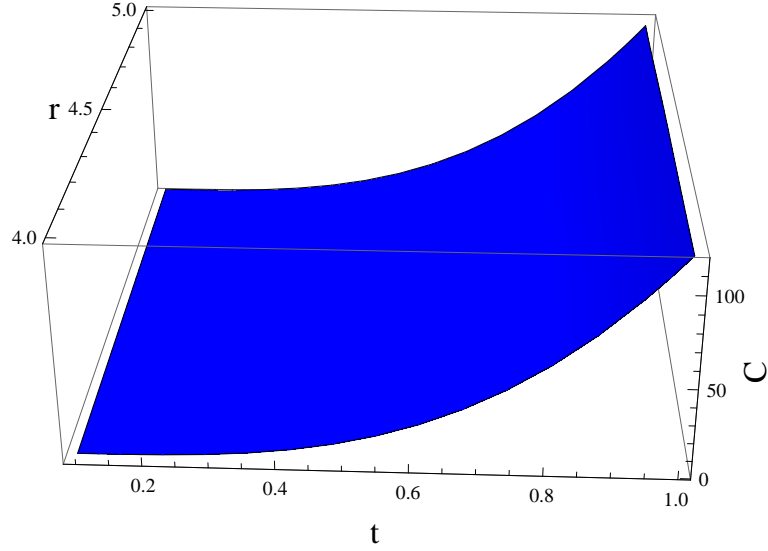,width=0.56\linewidth} \caption{Plots of $C$
versus $t$ and $r$ for $\kappa=-0.5$ and $m=1.4$, indicating the open universe increasing exponentially with the passage of time
.}
\end{figure}

\begin{figure}
\epsfig{file=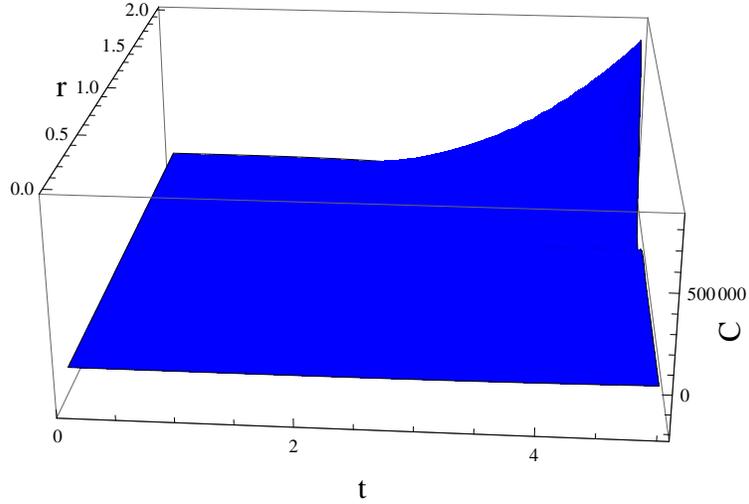,width=0.56\linewidth} \caption{Plots of $C$
versus $t$ and $r$ for $\kappa=0$ and $m=1.4$, represents flat universe .}
\end{figure}

\begin{figure}
\epsfig{file=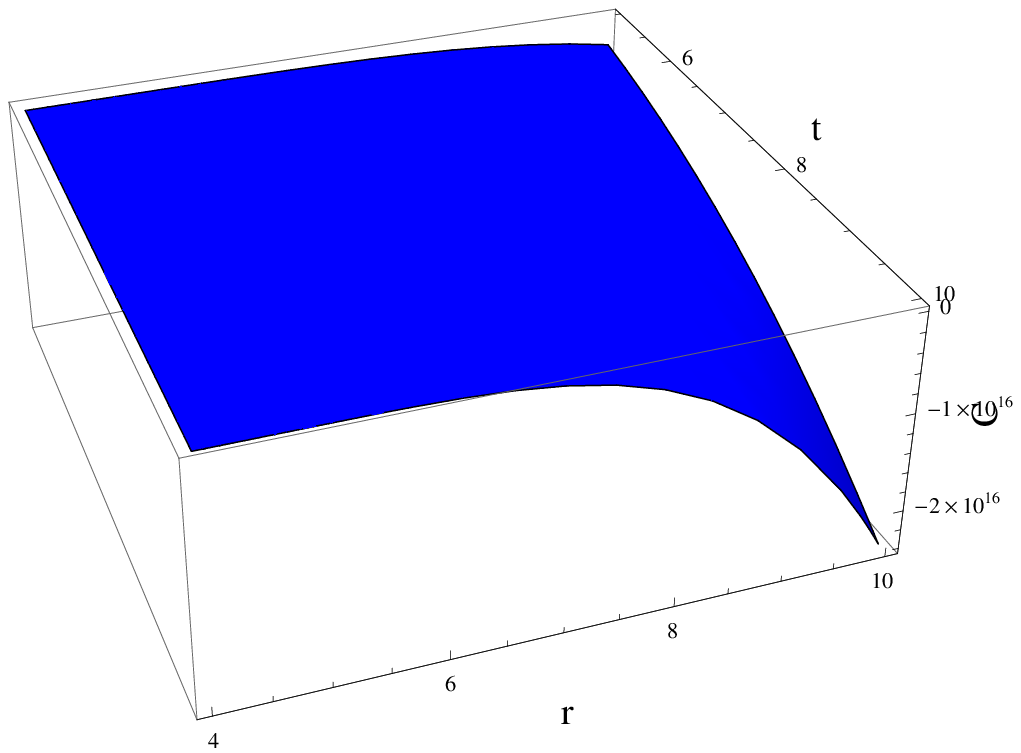,width=0.56\linewidth} \caption{Plots of $C$
versus $t$ and $r$ for $\kappa=0.5$ and $m=1.4$, indicating that $ f(R) $ terms effects the close universe instead of expanding to maximum size and then collapsing, it expands exponentially..}
\end{figure}
Substituting the above expression into \eqref{8a}, we get the LTB model in the form
\begin{align}\label{65a}
ds^2=-dt^2+\frac{C'^2}{1+\kappa(r)} dr^2+C^2 (d\theta^2+sin^2\theta d\phi^2).
\end{align}
The metric \eqref{65a} commonly linked with an inhomogeneous dust,
but it is worth noticing that an anisotropic fluid is the most
general source consistent with LTB spacetime. Now, Bianchi identity
reduces the following form in the case of non-dissipative dust
\begin{align}\label{66a}
\dot{\mu}_{eff}=\mu_{eff} \left(\frac{\dot{B}}{B}+2\frac{\dot{C}}{C}+\frac{5\dot{F}}{2F}\right).
\end{align}
Integrating Eq.\eqref{66a}, we obtain
\begin{align}\label{67a}
\mu_{eff}=\frac{3 h(r)(1+\kappa(r))^{\frac{1}{2}}}{(C^3)' F^{\frac{5}{2}}},
\end{align}
where $ h(r) $ is an integration function. Now, the structure
scalars $ Y_{T},~Y_{T F},~X_{T F},$ and $X_{T} $ for the line
element \eqref{65a} take the following form
\begin{align}\label{68a}
Y_{T}=&\frac{\ddot{C}}{C}-\frac{\ddot{C'}}{C'}-\frac{\chi_{22}}{C^2}+M^{(D)}_{1},
\\\label{69a}
 Y_{T F}=&-2\frac{\ddot{C}}{C}-\frac{\ddot{C'}}{C'}-\frac{\chi_{22}}{C^2}+\frac{4\pi}{F} \chi_{11}+M^{(D)}_{2}
\\\label{70a}
X_{T}=&\frac{2\dot{C} \dot{C'}}{C C'}+\frac{\dot{C}^2}{C^2}-\frac{\kappa}{C^2}-\frac{\kappa'}{C C'}-\frac{8\pi}{F} \chi_{00}+M^{(D)}_{3},
\\\label{71a}
X_{T F}=&\frac{\dot{C}^2}{C^2}-\frac{\dot{C} \dot{C'}}{C C'}-\frac{\kappa}{C^2}+\frac{\kappa'}{2C' C}-\frac{\chi_{22}}{2C^2}+\frac{4\pi}{F} \chi_{11}+M^{(D)}_{4}.
\end{align}
From Eqs.\eqref{68a} and \eqref{69a}, we have seen that the scalar
functions $ Y_{TF} $ and $ Y_{T} $ do not include the term $ \kappa
$ in their expressions.

\section{Generalization of LTB for dissipative flux}

In this section, we generalize the LTB spacetime named as GLTB by including the heat flux and obtain structure scalars for GLTB spacetime. We compare the structure scalars corresponding to LTB and GLTB to gain similarities between them while on the other hand, we discuss generalization based on symmetry using two cases namely diffusion and streaming out limit. To obtain the temperature profile for diffusion approximation, we presented a study of the transport equation. We have found from our preceding study that LTB spacetime ruled out the dissipative fluxes. Therefore, the main consequence of extending LTB spacetime is to generalize it for dissipative fluxes in the presence of dark source terms. For a specific situation, when dissipative fluxes vanish then GLTB reduces to LTB spacetime. We assume the geodesic dust fluid to find the spacetimes close to LTB. It is important to mention that the pure dust state represents the fluid to be geodesic in the absence of dissipation. Also, the generalization of LTB has a different form
as compared to that of GR and Palatini $ f(R) $ gravity.
Now, integration of Eq.\eqref{12a} produces
\begin{align}\label{72a}
B(t,r)=\frac{C'}{(1+K(t,r))^{\frac{1}{2}}},
\end{align}
where $ 1+K(t,r)=\left[C(r)+\int 4\pi \tilde{q}_{eff} R dt\right]^2 $ and
$ \tilde{q}_{eff} $ is defined as
$ \tilde{q}_{eff}=\frac{1}{F}\left(\tilde{q}+\frac{\chi_{01}}{B}\right) $ while the value
of $ \chi_{01} $ in case of Palatini formalism is given by
\begin{align}\nonumber
\chi_{01}=&\frac{1}{\kappa}
\left(\dot{F'}-\frac{5\dot{F}F'}{2F}-\frac{\dot{B}F'}{B}\right).
\end{align}
In case of GR, $ F \longrightarrow 1 $ and $ \chi_{01}
\longrightarrow 0 $ and $  \tilde{q}_{eff} $ becomes $ \tilde{q} $.
However, in case of metric $ f(R) $ formalism, the term $
\frac{5\dot{F}F'}{2F} $  in above expression is absent. Therefore, the
generalized LTB metric in case of Palatini and metric formalism is
given by
\begin{align}\label{73a}
ds^2=-dt^2+\frac{ (C')^2}{\left[C(r)+\int 4\pi \tilde{q}_{eff} C
dt\right]^2} dr^2+C^2(d\theta^2+sin^2\theta d\phi^2),
\end{align}
where, the value of $ \tilde{q}_{eff} $ is different for both
formalism and in GR, $ \tilde{q}_{eff} $ reduces to $ \tilde{q} $.
The generalized metric in Eq.\eqref{73a} is obtained by substituting
Eq.\eqref{72a} in the general line element given in Eq.\eqref{8a}.
The scalar functions $ Y_{T},~Y_{T F},~X_{T},$ and $ X_{T F} $ for
the line element \eqref{73a} are
\begin{align}\label{74a}
Y_{T}=&-2\frac{\ddot{C}}{C}-\frac{\ddot{C'}}{C'}+\frac{\dot{K}}{1+K}\left(\frac{\dot{C'}}{C'}-\frac{3\dot{K}}{4(1+K)}\right)+\frac{\ddot{K}}{2(1+K)}
-\frac{\chi_{22}}{C^2}+M^{(D)}_{1},
\\\label{75a}
Y_{T F}=&\frac{\ddot{C}}{C}-\frac{\ddot{C'}}{C'}+\frac{\dot{K}}{1+K} \left(\frac{\dot{C'}}{C'}-\frac{3\dot{K}}{4(1+K)}\right)
+\frac{\ddot{K}}{2(1+K)}-\frac{\chi_{22}}{C^2}+\frac{4\pi}{F} \chi_{11}+M^{(D)}_{2},
\\\label{76a}
X_{T}=&\frac{2\dot{C} \dot{C'}}{C C'}+\frac{\dot{C}^2}{C^2}-\frac{K}{C^2}-\frac{K'}{C C'}-\frac{\dot{C} \dot{K}}{C (1+K)}-\frac{8\pi}{F} \chi_{00}+M^{(D)}_{3},
\\\label{77a}
X_{T F}=&\frac{\dot{C}^2}{C^2}-\frac{\dot{C} \dot{C'}}{C C'}-\frac{K}{C^2}+\frac{K'}{2C C'}+\frac{\dot{C} \dot{K}}{2C(1+K)}-\frac{\chi_{22}}{2C^2}
+\frac{4\pi}{F} \chi_{11}+M^{(D)}_{4}.
\end{align}
Here, we need to enforce some conditions to continue further and to acquire some specific solutions. The selection of such conditions will be based on the measure that the obtained solutions are very close to LTB spacetime. In this framework, we will consider extensions of LTB spacetime based on structure scalars. On the other hand, we will discuss extensions in the context of symmetry properties. But before that, we explain an overview of the transport equation in the pure diffusion case.

\subsection{Extension via structure scalars}

From the study of structure scalars, we have concluded the following
remarks. The evolution of shear and expansion scalar is completely
controlled by structure scalars namely $ Y_{T}$ and $ Y_{T F} $ in
the case of geodesic fluid whether it is dissipative or not. The
difference between their terms both in LTB and GLTB are the same.
The term $ \kappa $ is not present in the expression of $ Y_{T}$ and
$ Y_{T F} $ in case of LTB spacetimes. Now, we consider the
structure scalars $ Y_{T}$ and $ Y_{T F} $ in such a way that in
both cases they contribute their same expression on the view of
above mentioned remarks and on the condition that maximal similarity
exist between LTB and GLTB. Therefore, by comparing Eqs.\eqref{69a}
and \eqref{75a}, we have
\begin{align}\label{78a}
\frac{\ddot{K}}{2(1+K)}+\frac{\dot{K}}{1+K} \left[-\frac{3\dot{K}}{4(1+K)}+\frac{\dot{C'}}{C'}\right]=0.
\end{align}
Integration of Eq.\eqref{78a} yields
\begin{align}\label{79a}
\frac{C' \dot{K}^{\frac{1}{2}}}{(1+K)^{\frac{3}{4}}}=C_{1}(r),
\end{align}
where $ C_{1}(r) $ is an integration  function. Using Eq.\eqref{72a}
in \eqref{79a}, we get
\begin{align}\label{80a}
C_{1}(r)=\frac{(8\pi \tilde{q}_{eff} C)^{\frac{1}{2}} C'}{(1+\kappa(r))^{\frac{1}{2}}+\int 4\pi \tilde{q}_{eff} C dt}.
\end{align}
Integration of Eq.\eqref{79a} produces
\begin{align}\label{81a}
K+1=\frac{4}{ \left[C_{1}(r)^2 \int \frac{dt}{C'^2}+C_{2}(r)\right]^2},
\end{align}
where $ C_{2}(r) $ is another arbitrary integration function.
Differentiation of Eq.\eqref{81a} with respect to $ t$ gives
\begin{align}\label{82a}
\dot{K}=- \frac{8C_{1}^2(r)}{(C')^2 \left[C_{2}(r)+C_{1}^2(r) \int \frac{dt}{C'^2}\right]^3}.
\end{align}
By comparing Eqs.\eqref{72a} and \eqref{82a}, we get
\begin{align}\label{83a}
2\pi \tilde{q}_{eff}=\frac{C_{1}^2(r)}{C (C')^2 \left[C_{2}(r)+C_{1}^2(r) \int \frac{dt}{(C')^2}\right]^2}.
\end{align}
Now, comparison of Eqs.\eqref{80a} and \eqref{83a} give the
relationship between integration functions $ C(r) $ and $ C_{2}(r) $
as follows
\begin{align}\label{84a}
C_{2}(r)=\frac{2(1-4\pi \tilde{q}_{eff} C C'^2)}{C(r)+\int 4\pi \tilde{q}_{eff} C dt},
\end{align}
we can get a GLTB on the basis of the suppositions imposed on $
Y_{T} $ and $ Y_{TF} $. The dissipative flux of GLTB can be obtained
up to two functions of $ r $ by taking into account the radial and
temporal dependence of $ R$ from Eq.\eqref{82a}. The function $
C_{1}(r) $ must satisfy the regularity condition $ C_{1}(0) = 0$,
and its disappearance returns the LTB solution to its initial state.
The field equations provide the rest of the physical variables. Now,
we use an example from the parabolic subclass of LTB to demonstrate
the method above. As a result, we choose
\begin{align}\label{85a}
C(t,r)=f(r) (-t+T(r))^{\frac{2}{3}},
\end{align}
where $ f(r) $ and $ T(r) $ are the parameters for any parabolic LTB
solution. Utilizing Eq.\eqref{85a} we express the integral $ \int
\frac{dt}{(C')^2} $ in the following form
\begin{align}\label{86a}
I=\int \frac{dt}{(C')^2}=\int \frac{(T(r)-t)^{\frac{2}{3}}dt}{[f'(r)(T(r)-t)+\frac{2}{3} f(r) T'(r)]^2}.
\end{align}
Putting $ T(r)-t=\tau^3 $ in Eq.\eqref{86a} and integrating we get:
\begin{align}\nonumber
I=-3\int\frac{\tau^4d\tau}{(a\tau^{3}+b)^{2}}
\end{align}
\begin{align}\label{87a}
I=\frac{\tau^2}{a^{2}\tau^{3}+ab}+\frac{1}{3\sqrt[3]{ba^{5}}}\left[2\sqrt{3}\arctan\left(\frac{1-2\sqrt[3]{\frac{a}{b}\tau}}{\sqrt{3}}\right)
-\ln\left(\frac{a\tau^{3}+b}{(\sqrt[3]{a}\tau+\sqrt[3]{b})^{3}}\right)\right],
\end{align}
here $ b=\frac{2}{3}T'(r)f(r) $ and $ a=f'(r) $. Plugging
Eq.\eqref{87a} in \eqref{83a} and using Eq.\eqref{85a}, we have
\begin{align}\label{88a}
2\pi\tilde{q}_{eff}=\frac{C^{2}_{1}(r)f(r)^{-1}}{\left(C^{2}_{1}(r)\left[\frac{\tau^{2}}{a}+\frac{a\tau^{3}+b}{3\sqrt[3]{ba^{5}}}
\left[2\sqrt{3}\arctan\left(\frac{1-2\sqrt[3]{\frac{a}{b}\tau}}{\sqrt{3}}\right)-\ln\left(\frac{a\tau^{3}+b}{(\sqrt[3]{a}\tau+\sqrt[3]{b})^{3}}\right)\right]
\right]+C_{2}(r)(a\tau^{3}+b)\right)^{2}}.
\end{align}
The magnitude of dissipation is being controlled by the function$ C_{1}(r) $ and can be set as small as desired when we require perturbations on LTB. It is obvious from the results that this GLTB may turn into LTB in the absence of dissipation. The physical properties of GLTB have reasonable consequences for a smaller value of $ C_{1}(r) $ corresponding to LTB.

\subsection{Extension via symmetry properties}

We will discuss another method for obtaining GLTB spacetime in this subsection. It entails that any GLTB spacetime has a similar property as the corresponding LTB. The Lie derivative of energy momentum tensor with respect to vector field $ \xi $ is defined as
\begin{align}\label{89a}
{\L}_{\xi} T_{\zeta\varrho}=0,
\end{align}
Then, using the effective energy-momentum tensor with dissipation in Eq.\eqref{104a}, we obtain the following equations
\begin{align}\label{90a}
&\xi^{0} \dot{\tilde{\mu}}_{eff}+\xi^{1} \tilde{\mu}'_{eff}+2\tilde{\mu}_{eff} \xi^{0}_{,0}-2\tilde{q}_{eff} B \xi^{1}_{,0}=0,
\\\label{91a}
&-\xi^{0} \dot{\tilde{q}}_{eff}-\xi^{1} \tilde{q}'_{eff}+\tilde{\mu}_{eff} \frac{\xi^{0}_{,1}}{B}-\tilde{q}_{eff}\left(\xi^{0}_{,0}+\xi^{0}\frac{\dot{B}}{B}+\xi^{1}\frac{B'}{B}+\xi^{1}_{,1}\right)+\epsilon_{eff} B \xi^{1}_{,0}=0,
\\\label{92a}
&\xi^0 \dot{\epsilon}_{eff}+\xi^1 \epsilon'_{eff}-2\tilde{q}_{eff} \frac{\xi^0_{,1}}{B}+2\epsilon_{eff} \left(\xi^0 \frac{\dot{B}}{B}+\xi^1 \frac{B'}{B}+\xi^1{,1} \right)=0.
\end{align}
In order to generalize LTB on the basis of symmetry in the presence of Palatini correction terms, we will discuss two techniques.
Diffusion approximation $ ( q_{eff}\neq0,~\epsilon_{eff}=0,~\tilde{q}_{eff}=q_{eff}) $ and streaming out limit $ (\epsilon_{eff}\neq0,~q_{eff}=0, \tilde{q}_{eff}=\epsilon_{eff})$.

\subsection{Diffusion approximation}

In case of heat flux $ ( q_{eff}\neq0, \epsilon_{eff}=0,
\tilde{q}_{eff}=q_{eff}) $ using Eqs.\eqref{12a} and \eqref{72a}, we
obtain
\begin{align}\label{93a}
8\pi q_{eff} B=\frac{\dot{K}}{1+K} \frac{C'}{C},
\end{align}
in this case, we get from Eq.\eqref{45a}
\begin{align}\label{94a}
q_{eff} =\frac{g(r)D_{3}}{B^2 C^2 F^3},
\end{align}
where $ D_{3}=exp\left[-\int\frac{F'}{2F B q_{eff}} \mu_{eff}
dt\right]$, and $ g(r) $ serves as an arbitrary function.
 Combining Eqs.\eqref{72a},
\eqref{93a} and \eqref{94a}, the following result is obtained as
\begin{align}\label{95a}
8\pi g(r)= \frac{C C'^2 F^{3}}{D_{3}} \frac{\dot{K}}{(1+K)^{\frac{3}{2}}}.
\end{align}
Further, Eq.\eqref{92a} implies that $ \xi^0 =F(t) $ in case of LTB,
and using this result, Eqs.\eqref{90a} and \eqref{91a} becomes
\begin{align}\label{96a}
F(t) \dot{\mu}_{eff}+\xi^1 \mu'_{eff}+2\mu_{eff} \dot{F}(t) -2q_{eff} B \xi^1_{,0}=0,
\\\label{97a}
F(t) \dot{q}_{eff}+\xi^1 q'_{eff}+q_{eff} \left(\dot{F}(t)+F(t)\frac{\dot{B}}{B}+\xi^1 \frac{B'}{B}+\xi^1_{,1}\right)=0.
\end{align}
Eq.\eqref{97a} can be written as
\begin{align}\label{98a}
\xi^1\left[\ln(q_{eff} B \xi^1)\right]'+F(t)  \left[\ln(q_{eff} B F(t))\right]^{.} =0.
\end{align}
Multiplying Eq.\eqref{98a} by $ q_{eff}B $, we obtain
\begin{align}\label{99a}
(q_{eff} B \xi^1)'+(q_{eff} F B)^{.}=0,
\end{align}
Now it is possible to write the partial solution of Eq.\eqref{99a}
as
\begin{align}\label{100a}
q_{eff} B \xi^1 =-\dot{\psi}(t,r),
\\\label{101a}
q_{eff} B F(t)=\psi'(t,r),
\end{align}
using Eqs.\eqref{93a} and \eqref{101a}, we get
\begin{align}\label{102a}
\psi'(t,r)=\frac{F(t)}{8\pi}\frac{C'}{C}\frac{\dot{K}}{1+K}.
\end{align}
Now, manipulating Eqs.\eqref{72a}, \eqref{94a}, \eqref{101a} and
\eqref{102a}, we have
\begin{align}\label{103a}
8\pi g(r)=\frac{C C'^2 F^{3}}{D_{3}}
\frac{\dot{K}}{(1+K)^{\frac{3}{2}}},
\end{align}
which is accurately match with Eq.\eqref{95a}. Now, integrate
Eq.\eqref{95a} to get
\begin{align}\label{104a}
\frac{1}{(1+K)^{\frac{1}{2}}}= - 4\pi g(r) \int \frac{D_{3}dt}{C C'^2 F^3}+C'(r),
\end{align}
where $ C'(r) $ is an integration function. For a particular case,
using the form of $ C(t,r) $ given in Eq.\eqref{85a}, we get
\begin{align}\nonumber
\frac{1}{(1+K)^{\frac{1}{2}}}=&\frac{-4\pi g(r)}{f(r) f'(r)[f'(r)(T(r)-t)+\frac{2}{3} f(r) T'(r)]F^3}+C'(r)+4\pi g(r)
\\\label{105a}
&\int \frac{3\dot{F} F^2}{[f'(r)(T(r)-t)+\frac{2}{3}f(r) T'(r)]}.
\end{align}
Inserting Eq.\eqref{105a} in \eqref{94a} and using Eq.\eqref{72a},
we obtain
\begin{align}\nonumber
q_{eff}=&\frac{g(r)}{F^{3}f^{2}(r)(T(r)-t)^{\frac{2}{3}}}\left[-\frac{4\pi
g(r)}{f(r)f'(r)F^{3}}+\left(4\pi g(r)\int \frac{3\dot{F}
F^2}{(f'(r)(T(r)-t)+\frac{2}{3}f(r)
T'(r))}\right.\right.\\\label{106a}
&\left.\left.+C'(r)\right)
\left(f'(r)(T(r)-t)+\frac{2}{3} f(r) T'(r)\right)\right]^{-2}.
\end{align}
Therefore, the GLTB has been characterized by four functions, but due to the invariance of radial coordinate, one function vanishes. A couple of functions may be obtained from the LTB seed model and the rest of them is found from the equations under the hypothesis that LTB solution has similar radial dependence as $\xi^{1}$. Furthermore, the temperature profile for the diffusion case may be obtained from the transport equation.

\subsection{Streaming-out approximation }

In the case of null radiation, $ (q_{eff}=0,~\epsilon_{eff}\neq0) $, the Bianchi identities reduces in the form
\begin{align}\label{107a}
\dot{\mu}_{eff}+\mu_{eff}\left(\frac{\dot{B}}{B}+2\frac{\dot{C}}{C}+\frac{5\dot{F}}{2F}-\frac{F'}{2F B}\right)=0.
\end{align}
Integration of Eq.\eqref{107a} yields
\begin{align}\label{108a}
\mu_{eff}=\frac{j(r)}{B C^2 F^{\frac{5}{2}}} \exp \int \frac{F' dt}{2F B}.
\end{align}
Using Eq.\eqref{72a} in \eqref{108a}, we get
\begin{align}\label{109a}
\mu_{eff}=\frac{3j(r)[\int 4\pi \epsilon_{eff}Cdt+(1+\kappa(r))^{\frac{1}{2}}]}{(C^{3})'F^{\frac{5}{2}}} \exp \int \frac{F' dt}{2F B},
\end{align}
where $ j(r) $ is constant of integration. Now, under the assumption
$ (q_{eff}=0,~\epsilon_{eff}\neq0) $, Eqs.\eqref{90a}-\eqref{92a}
take the following form
\begin{align}\label{110a}
&\xi^0(\mu_{eff}+\epsilon_{eff})^{.}+\xi^1 (\mu_{eff}+\epsilon_{eff})'+2(\mu_{eff}+\epsilon_{eff})\xi^0_{,0}-2\epsilon_{eff} B \xi^1_{,0}=0,
\\\label{111a}
&\xi^0 \dot{\epsilon}_{eff}+\xi^1 \epsilon'_{eff}-(\mu_{eff}+\epsilon_{eff})\frac{\xi^0_{,1}}{B}+\epsilon_{eff} \left(\xi^0_{,0}+\xi^0 \frac{\dot{B}}{B}+\xi^1 \frac{B'}{B}+\xi^1_{,1}\right)-\epsilon_{eff} B \xi^1_{,0}=0,
\\\label{112a}
&\xi^0 \dot{\epsilon}_{eff}+\xi^1 \epsilon'_{eff}-2\epsilon_{eff} \frac{\xi^0_{,1}}{B}+2\epsilon_{eff}\left(\xi^0 \frac{\dot{B}}{B}+\xi^1 \frac{B'}{B}+\xi^1_{,1}\right)=0.
\end{align}
From Eqs.\eqref{110a}-\eqref{112a}, we obtain
\begin{align}\label{113a}
\xi^0 \dot{\mu}_{eff}+\xi^1 \mu'_{eff}+2 \mu_{eff} \xi^0_{,0}+2 \mu_{eff} \frac{\xi^0_{,1}}{B}=0.
\end{align}
After some manipulations, Eq.\eqref{113a} gives
\begin{align}\label{114a}
\xi^1=-\frac{\mu_{eff}}{\mu'_{eff}} \left(2\dot{F}(r,t)+F(r,t) \frac{\dot{\mu}_{eff}}{\mu_{eff}}+\frac{2 F'(r,t)}{B}\right),
\end{align}
where $ \xi^0=F(r,t)$. Suppose that $ \xi^0=F(t), $ in order to
satisfy the obligation of maximal resemblance between LTB and GLTB.
In that case, Eqs.\eqref{110a}-\eqref{112a} implies
\begin{align}\label{115a}
&F(t)(\mu_{eff}+\epsilon_{eff})^{.}+\xi^1 (\mu_{eff}+\epsilon_{eff})'+2 (\mu_{eff}+\epsilon_{eff}) \dot{F}(t)-2\epsilon_{eff} B \xi^1_{,0}=0,
\\\label{116a}
&F(t) \dot{\epsilon}_{eff}+\xi^1 \epsilon'_{eff}+\epsilon_{eff} \left(\dot{F}(t)+F(t)\frac{\dot{B}}{B}+\xi^1 \frac{B'}{B}+\xi^1_{,1} \right)-\epsilon_{eff} B \xi^1_{,0}=0,
\\\label{117a}
&F(t) \dot{\epsilon}_{eff}+\xi^1 \epsilon'_{eff}+2\epsilon_{eff} \left(F(t) \frac{\dot{B}}{B}+\xi^1 \frac{B'}{B}+\xi^1_{,1} \right)=0.
\end{align}
Eq.\eqref{117a} may turn into the following form
\begin{align}\label{118a}
F(t) \left[\ln(\epsilon_{eff} B^2 )\right]^{.} + \xi^1 \left[\ln(\epsilon_{eff} (B \xi^1)^2 ) \right]'=0,
\end{align}
or it may be rewritten as
\begin{align}\label{119a}
F(t) (\epsilon_{eff} B^2)^{.} +\frac{1}{\xi^1} (\epsilon_{eff} (B \xi^1)^2 )'=0.
\end{align}
Also, in the presence of these conditions, Eq.\eqref{45a} gets the
following form
\begin{align}\label{120a}
\left[\ln(\epsilon_{eff}(B F C)^2 ) \right]^{.} +\frac{1}{B} \left[\ln(\epsilon_{eff} C^2) \right]'+\frac{\dot{F}}{F}+\frac{3 F'}{B}
+\frac{F'}{2F B} \left(\frac{\mu_{eff}}{\epsilon_{eff}}\right)=0,
\end{align}
or may be written as
\begin{align}\label{121a}
\left[\epsilon_{eff} (B F C)^2 \right]^{.}+B F^2 (\epsilon_{eff} C^2)'+B^2 F \dot{F} C^2 \epsilon_{eff}+\frac{F' B C^2 F}{2} \mu_{eff}+3B F^2 F' C^2 \epsilon_{eff}=0.
\end{align}
It is now possible to acquire GLTB based on LTB (in the streaming
out approximation) in the presence of Palatini $ f(R) $ gravity by
adopting the following procedure. Suppose that LTB solution has the
similar type of radial dependence as $ \xi^{1} $. Obtain the value
of $ (\epsilon_{eff}B^{2}) $ from Eq.\eqref{121a} and then
substitute in Eq.\eqref{119a}. Taking into account a definite LTB
model with a given field $ \xi $ and take $ F(t)=1 $, because it is
arbitrary. Integrating Eq.\eqref{119a} to get the effective
radiation density with respect to $ r $.

\section{Discussion}

It is well known that at large scales, the cold dark matter is non-collisional and
have strong influence of rest-mass, where the dissipative and pressure terms have negligible
influence due to their kinetic nature. Therefore, it would be significant to examine the stability of such analysis
which have influence of some kinetic terms, e.g., dissipative fluxes within the background of a particular modified gravity namely,
Palatini $f(R)$ gravity.
It is important to mention that the state of pure dust suggests that the fluid
is geodesic in the absence of dissipation, while this condition does not hold when we consider dissipative flux. Herrera \emph{et al}. \cite{herrera2015physical} considered the dissipative dust fluid as a source of gravitational radiation and have mentioned
that the dissipative dust fluid is the most compatible fluid to discuss the gravitational
radiations. Sharif and Yousaf
\cite{sharif2015stability} analyzed the factors that generate the
energy density irregularities in expansion free matter distribution
by considering the dissipative and non-dissipative case in Palatini
$ f(R) $ gravity. They considered the particular case of dissipative
dust fluid and apply certain conditions to examine the contribution
of radiating parameters in the origon of energy density
irregularities.

Herrera \emph{et al}. \cite{herrera2010lemaitre}
discussed the generalization of LTB spacetime with the help of
structure scalars and symmetric properties. They considered the
dissipative dust fluid to generalize the LTB spacetime for the
dissipative case because LTB does agree with dissipative phenomena.
They have mentioned that the state of pure dust suggests that the fluid is geodesic in the absence of dissipation while this condition remains not valid when we consider dissipative flux.
Yousaf \emph{et al}. \cite{yousaf2016causes} considered the imperfect fluid to examine
the factors that create irregularities for a spherical star in $ f(R,T) $ gravity.
They studied particular cases of anisotropic, isotropic, and dust to analyze the
irregularity factors in the dissipative and non-dissipative scenarios. They considered
dissipative dust as a special case to analyze the effects of dissipation in the form of
heat flux and null radiations. Herrera \cite{herrera2015gravitational} studied the
gravitational radiation and its properties by considering dissipative fluids as well as
addressed some particular cases such as shear-free and perfect fluid to analyze
their behavior. They found that dust fluid along with dissipation is the more consistent
fluid distribution with gravitational radiation.

In this manuscript, we have considered the effects of Palatini $ f(R)$ terms on the generalization of LTB spacetime. We start our analysis by taking into account the spherically symmetric distribution of dust fluid. Our main purpose is to analyze the contribution of dark source terms obtained from the modification of Einstein-Hilbert's action. To do so, we have evaluated the modified field equations and kinematical variables associated with fluid in the framework of $ f(R) $ gravity. We have found that the presence of four acceleration is totally dependent on Palatini $ f(R)$ formalism as observed from Eq.\eqref{17a}. We have used the expression for mass function introduced by Misner and Sharp \cite{misner1964relativistic} to observe the quantity of matter inside the geometrical object. The junction conditions are obtained for the smooth matching of exterior and interior geometry. A relation between Misner-Sharp mass, Weyl scalar and fluid variables along with dark source terms has been established to determine the gravitational contribution. To examine the dynamics of stellar bodies, we have calculated Bianchi identities, the Raychaudhuri equation, and a differential equation for Weyl scalar.
We introduce the tensors $ Y_{\zeta\varrho} $ and $ X_{\zeta\varrho} $ obtained from the orthogonal decomposition method proposed by Herrera \cite{herrera2009structure} and found their expressions in the framework of Palatini $ f(R) $ gravity. We split the tensors $Y_{\zeta\varrho}$ and $X_{\zeta\varrho}$ into their trace and trace-free parts. These scalar functions are named as Palatini $ f(R) $ structure scalars and their physical significance have been discussed in the literature. It is worthwhile to mention that the scalar $ Y_{T} $ gives information about pressure anisotropy in the presence of extra curvature terms due to $ f(R) $ gravity and the existence of an unknown form of energy and matter is indicated by these extra curvature terms. The scalar $X_{T}$ addresses the energy density of the fluid and few extra terms describing the effects of DE and DM. The scalar $Y_{TF}$ comprises pressure anisotropy with the addition of the Weyl scalar and extra curvature terms. The already obtained evolution equations and differential equation for Weyl scalar have been transformed in terms of Palatini $ f(R) $ structure scalars. We obtained the Lema\^{\i}tre-Tolman-Bondi metric by integrating the field equation corresponding to the $ G_{01} $ component and discussed different solutions of the evolution equation depending on the value of $ \kappa $ in the case of LTB spacetime. The scalar functions for LTB spacetime have been obtained with extra curvature terms and generalize LTB spacetime by considering the heat fluxes and found structure scalars for GLTB spacetime. To obtain the maximal similarity between LTB and GLTB, we compare the structure scalars obtained from LTB and GLTB. We discussed the generalization based on symmetry using two techniques namely diffusion approximation and streaming-out limit, and obtained a temperature profile using the transport equation for diffusion case in the background of Palatini $ f(R) $ gravity.

\textbf{Appendix A}

The extra curvature terms appearing in the field equations are given as
\begin{align}\label{122a}
\chi_{00}=\frac{1}{\kappa} \left(\frac{-3\dot{F^2}}{2F}+\frac{F'^2}{2F B}+\frac{F''}{B}-\frac{\dot{B}\dot{F}}{B}-\frac{B' F'}{B^3}-\frac{2\dot{C}\dot{F}}{C}+\frac{2C' F'}{C B^2}+\frac{F R}{2}-\frac{f}{2}-\frac{3F'^2}{4F B^2}-\frac{9\dot{F}^2}{4F}\right),
\end{align}
\begin{align}\label{123a}
\chi_{11}=&\frac{1}{\kappa B^2} \left(\frac{B^2 \dot{F}^2}{2F}+B^2 \ddot{F}-\frac{3F'^2}{2F}+\frac{2B^2 \dot{C}\dot{F}}{C}-\frac{2C' F'}{C}-\frac{B^2 R F}{2} -\frac{f B^2}{2}-\frac{3B^2 \dot{F}^2}{4F}-\frac{3F'^2}{4F}\right),
\\\label{124a}
\chi_{01}=&\frac{1}{\kappa} \left(\dot{F'}-\frac{\dot{B}F'}{B}-\frac{5\dot{F}F'}{2F}\right),
\\\nonumber
\chi_{22}=&\frac{8\pi}{F \kappa} \left(C \dot{C}\dot{F}+\frac{C^2
\dot{F}^2}{2F}-\frac{C C' F'}{B^2}-\frac{C^2 F'^2}{2F B^2}+C^2
\ddot{F}-\frac{C^2 F''}{B^2}+\frac{C^2 \dot{B}\dot{F}}{B}+\frac{C^2
B' F'}{B^3}-\frac{F C^2 R}{2}\right.
\\\label{125a}
&\left.-\frac{f C^2}{2}-\frac{3C^2 \dot{F}^2}{4F}+\frac{3C^2 F'^2}{4F B^2}\right).
\end{align}

\textbf{Appendix B}

The extra curvature terms appearing in the structure scalars are given as
\begin{align}\label{126a}
M^{(D)}_{1}=&\frac{1}{F} \left(\ddot{F}+\frac{\dot{F}^2}{2 F}+\frac{F'^2}{4 F B^2}+\frac{\dot{C} \dot{F}}{C}-\frac{C' F'}{C B^2}-\frac{F''}{B^2}+\frac{\dot{B} \dot{F}}{B}+\frac{B' F'}{B^3}+\frac{(f-F R)}{2}\right),
\\\nonumber
M^{(D)}_{\zeta\varrho}=&-\frac{1}{2 F} \left[(\nabla_{\zeta} \nabla_{\varrho} F-g_{\zeta\varrho} \Box F)-(\nabla_{\gamma} \nabla_{\delta} F-g_{\gamma\delta} \Box F)g_{\zeta\varrho} V^{\gamma} V^{\delta} \right]-\frac{3}{4F^2} \nabla_{\gamma} F \nabla_{\delta} F V^{\gamma}
\\\label{127a}
& V^{\delta} g_{\zeta\varrho}+\frac{1}{F}\left[\left(\frac{1}{F}\nabla_{\gamma} F \nabla_{\delta} F-\frac{2}{3}(\nabla_{\gamma} \nabla_{\delta} F-g_{\gamma\delta} \Box F)\right) V^{\gamma} V^{\delta} h_{\zeta\varrho}\right]-\frac{1}{2F} \Box F h_{\zeta\varrho},
\\\nonumber
M^{(D)}_{3}=&\frac{1}{8F} \epsilon^{\epsilon\rho\varrho}\left[\epsilon_{\pi\rho\varrho}\left(\nabla^{\pi} \nabla_{\epsilon} F-\frac{3}{2F} \nabla^{\pi} F \nabla_{\epsilon} F\right)-\epsilon_{\pi\epsilon\varrho}\left(\nabla^{\pi} \nabla_{\rho} F-\frac{3}{2F} \nabla^{\pi} F \nabla_{\rho} F\right)-\epsilon_{\rho\theta\varrho}\right.
\\\nonumber
&\left.\left(\nabla^{\theta} \nabla_{\epsilon} F-\frac{3}{2F} \nabla^{\theta} F \nabla_{\epsilon} F\right)+\epsilon_{\epsilon\theta\varrho}\left(\nabla^{\theta} \nabla_{\rho} F-\frac{3}{2F} \nabla^{\theta} F \nabla_{\rho} F\right)\right]-\frac{1}{F} h^{\varrho}_{\varrho}\left(\frac{3}{4F}(\nabla F)^2\right.
\\\label{128a}
&\left.+\frac{(f-R F)}{2}-\Box F\right) \frac{1}{F}\left(3 \Box F+2(R F -f)-\frac{3}{F} (\nabla F)^2-\frac{3}{2F}g^{\zeta\varrho} \nabla_{\zeta} \nabla_{\varrho} F \right),
\\\nonumber
N^{(D)}_{\zeta\varrho}&=\frac{1}{8F}
\epsilon^{\epsilon\rho\xi}\left(g_{\xi\zeta}-\frac{1}{3}
h_{\xi\zeta}\right)\left[\epsilon_{\pi\rho\varrho}\left(\nabla^{\pi}
\nabla_{\epsilon} F-\frac{3}{2F} \nabla^{\pi} F \nabla_{\epsilon}
F\right)-\epsilon_{\pi\epsilon\varrho}\left(\nabla^{\pi}
\nabla_{\rho} F-\frac{3}{2F} \right.\right.
\\\nonumber
&\left.\left.\times \nabla^{\pi} F \nabla_{\rho}
F\right)-\epsilon_{\rho\theta\varrho}\left(\nabla^{\theta}
\nabla_{\epsilon} F-\frac{3}{2F} \nabla^{\theta} F \nabla_{\epsilon}
F\right)+\epsilon_{\epsilon\theta\varrho}\left(\nabla^{\theta}
\nabla_{\rho} F-\frac{3}{2F} \nabla^{\theta} F \nabla_{\rho}
F\right)\right]
\\\label{129a}
&-\frac{1}{F}\left(g_{\zeta\varrho}-\frac{1}{3} h_{\zeta\varrho}\right) \left(\frac{3}{4F}(\nabla F)^2+\frac{(f-R F)}{2}-\Box F \right).
\end{align}

\textbf{Appendix C}

The extra curvature terms appearing in the evolution equations of shear and expansion are given as
\begin{align}\label{130a}
D_{1}&=\frac{2\ddot{F}}{F}-\frac{7\dot{F}^2}{12F^2}+\frac{4\dot{B}\dot{F}}{3B
F}+\frac{8\dot{C}\dot{F}}{3C F}-\frac{F'^2}{8F^2 B^2}, \\\nonumber
D_{2}&=\frac{2}{3}\left[\left(\frac{\dot{B}}{B}-\frac{\dot{C}}{C}\right)+\frac{\dot{F}^2}{8F^2}-\frac{3F'^2}{16F^{2}B^{2}}\right]^{\frac{3}{2}}+
\frac{8\dot{F}}{3F}\left(\frac{\dot{B}}{B}-\frac{\dot{C}}{C}\right)+\frac{4}{3}\left(\frac{\dot{B}}{B}+2\frac{\dot{C}}{C}+2\frac{\dot{F}}{F}\right)
\\\label{131a}
&\left(\frac{\dot{F}^2}{8F^2}-\frac{3F'^2}{16F^2 B^2}\right)+\left(\frac{\dot{F}^2}{8F^2}\right)^{.}-\left(\frac{3F'^2}{16F^2 B^2}\right)^{.}.
\end{align}

\textbf{Appendix D}

\begin{align}\nonumber
Y_{\zeta\varrho}=&E_{\zeta\varrho}+\frac{\kappa}{2F}(\tilde{\mu}
h_{\zeta\varrho}-\epsilon \chi_{\zeta}
\chi_{\varrho})+\frac{\kappa}{3F}(-\tilde{\mu}+\epsilon)h_{\zeta\varrho}-\frac{1}{2F}\left[
(\nabla_{\zeta} \nabla_{\varrho} F -g_{\zeta\varrho}\Box
F)+(\nabla_{\gamma} \nabla_{\varrho} F \right.
\\\nonumber
&\left.-g_{\gamma\varrho} \Box F) V_{\zeta} V^{\gamma}+(\nabla_{\zeta} \nabla_{\delta} F-g_{\zeta\delta} \Box F) V_{\varrho} V^{\delta}-(\nabla_{\gamma} \nabla_{\delta} F -g_{\gamma\delta} \Box F) g_{\zeta\varrho} V^{\gamma} V^{\delta}\right]+\frac{3}{4F^2}
\\\nonumber
&\left[ \nabla_{\zeta} F \nabla_{\varrho} F +\nabla_{\gamma} F \nabla_{\varrho} F V_{\zeta} V^{\gamma}+\nabla_{\zeta} F \nabla_{\delta} F  V_{\varrho} V^{\delta} -\nabla_{\gamma} F \nabla_{\delta} F g_{\zeta\varrho} V^{\gamma} V^{\delta}\right]-\frac{1}{2F^2}
\\\label{132a}
&(g^{\zeta\varrho} \nabla_{\zeta} F \nabla_{\varrho} F-\Box F)h_{\zeta\varrho}+\frac{1}{6} (f-F R)h_{\zeta\varrho}+\frac{1}{4} (\nabla F)^2 h_{\zeta\varrho},
\\\nonumber
X_{\zeta\varrho}&=-E_{\zeta\varrho}+\frac{\kappa
\epsilon}{2F}(\chi_{\zeta}
\chi_{\varrho}-h_{\zeta\varrho})+\frac{\kappa}{3
F}(-\tilde{\mu}+\epsilon)h_{\zeta\varrho}+\frac{1}{8}
\epsilon^{\epsilon\rho}_{\zeta}\left[\epsilon_{\pi\rho\varrho}\left(\nabla^{\pi}
\nabla_{\epsilon} F-\frac{3}{2F} \nabla^{\pi} F \nabla_{\epsilon}
F\right)\right.
\\\nonumber
&\left.-\epsilon_{\pi\epsilon\varrho}\left(\nabla^{\pi} \nabla_{\rho} F-\frac{3}{2F} \nabla^{\pi} F \nabla_{\rho} F\right)-\epsilon_{\rho\theta\varrho}\left(\nabla^{\theta} \nabla_{\epsilon} F-\frac{3}{2F} \nabla^{\theta} F \nabla_{\epsilon} F\right)+\epsilon_{\epsilon\theta\varrho}\left(\nabla^{\theta} \nabla_{\rho} F \right.\right.
\\\label{133a}
&\left.\left.-\frac{3}{2F} \nabla^{\theta} F \nabla_{\rho} F\right)\right]-\frac{3}{2F^2} \nabla_{\zeta} F \nabla_{\varrho} F +\frac{1}{F}\left(\frac{1}{4 F} (\nabla F)^2+\frac{(f-R F)}{6}\right)
\end{align}


\vspace{0.3cm}


\begin{thebibliography}{10}

\bibitem{nojiri2005modified}
S.~Nojiri and S.~D. Odintsov {\em Phys. Lett. B}, vol.~631, no.~1-2, pp.~1--6,
  2005.

\bibitem{Yousaf2018}
Z.~Yousaf {\em Astrophys. Space Sci.}, vol.~363, no.~11, p.~226, 2018.

\bibitem{doi:10.1142/S021827181850044X}
M.~Z. Bhatti, M.~Sharif, Z.~Yousaf, and M.~Ilyas {\em Int. J. Mod. Phys. D},
  vol.~27, p.~1850044, 2018.

\bibitem{yousaf2019role}
Z.~Yousaf {\em Eur. Phys. J. Plus}, vol.~134, no.~5, p.~245, 2019.

\bibitem{houndjo2017higher}
M.~J.~S. Houndjo, M.~E. Rodrigues, N.~S. Mazhari, D.~Momeni, and R.~Myrzakulov
  {\em Int. J. Mod. Phys. D}, vol.~26, p.~1750024, 2017.

\bibitem{doi:10.1142/S0219887818501463}
Z.~Yousaf, M.~Sharif, M.~Ilyas, and M.~Z. Bhatti {\em Int. J. Geom. Meth. Mod.
  Phys.}, vol.~15, p.~1850146, 2018.

\bibitem{haghani2013further}
Z.~Haghani, T.~Harko, F.~S.~N. Lobo, H.~R. Sepangi, and S.~Shahidi {\em Phys.
  Rev. D}, vol.~88, no.~4, p.~044023, 2013.

\bibitem{odintsov2013f}
S.~D. Odintsov and D.~S{\'a}ez-G{\'o}mez {\em Phys. Lett. B}, vol.~725, no.~4,
  pp.~437--444, 2013.

\bibitem{yousaf2017stability}
Z.~Yousaf, M.~Z. Bhatti, and U.~Farwa {\em Class. Quantum Grav.}, vol.~34,
  p.~145002, 2017.

\bibitem{Yousaf2017}
Z.~Yousaf, M.~Z. Bhatti, and U.~Farwa {\em Eur. Phys. J. C}, vol.~77, p.~359,
  2017.

\bibitem{yousaf2016stability}
Z.~Yousaf, M.~Z. Bhatti, and U.~Farwa {\em Mon. Not. Roy. Astron. Soc.},
  p.~2698, 2016.

\bibitem{PhysRevD.72.083505}
G.~J. Olmo {\em Phys. Rev. D}, vol.~72, p.~083505, 2005.

\bibitem{doi:10.1142/S0218271811018925}
G.~J. Olmo {\em Int. J. Mod. Phys. D}, vol.~20, no.~04, pp.~413--462, 2011.

\bibitem{PhysRevD.86.044014}
G.~J. Olmo and D.~Rubiera-Garcia {\em Phys. Rev. D}, vol.~86, p.~044014, 2012.

\bibitem{PhysRevD.86.104039}
G.~J. Olmo, H.~Sanchis-Alepuz, and S.~Tripathi {\em Phys. Rev. D}, vol.~86,
  p.~104039, 2012.

\bibitem{PhysRevD.86.127504}
S.~Capozziello, T.~Harko, T.~S. Koivisto, F.~S.~N. Lobo, and G.~J. Olmo {\em
  Phys. Rev. D}, vol.~86, p.~127504, 2012.

\bibitem{olmo2020stellar}
G.~J. Olmo, D.~Rubiera-Garcia, and A.~Wojnar {\em Phys. Rep.}, vol.~876, p.~1,
  2020.

\bibitem{olmo2020junction}
G.~J. Olmo and D.~Rubiera-Garcia, ``Junction conditions in palatini $f
  (\textsc{R})$ gravity,'' {\em Classical and quantum gravity}, vol.~37,
  p.~215002, 2020.

\bibitem{sotiriou2006f}
T.~P. Sotiriou {\em Class. Quantum Grav.}, vol.~23, no.~17, p.~5117, 2006.

\bibitem{nojiri2007introduction}
S.~Nojiri and S.~D. Odintsov {\em Int. J. Geom. Meth. Mod. Phys.}, vol.~4,
  p.~115, 2007.

\bibitem{starobinsky2007disappearing}
A.~A. Starobinsky {\em JETP Lett.}, vol.~86, pp.~157--163, 2007.

\bibitem{santos2007energy}
J.~Santos, J.~S. Alcaniz, M.~J. Reboucas, and F.~C. Carvalho {\em Phys. Rev.
  D}, vol.~76, no.~8, p.~083513, 2007.

\bibitem{olmo2011hamiltonian}
G.~J. Olmo and H.~Sanchis-Alepuz, ``Hamiltonian formulation of palatini f (r)
  theories {\`a} la brans-dicke theory,'' {\em Phys. Rev. D .}, vol.~83,
  no.~10, p.~104036, 2011.

\bibitem{bamba2010thermodynamics}
K.~Bamba and C.~Geng, ``Thermodynamics in f (r) gravity in the palatini
  formalism j. cosmol,'' {\em Astropart. Phys. JCAP06 (2010)}, vol.~14,
  no.~1005.5234, 2010.

\bibitem{bhatti2019stability}
M.~Z. Bhatti, Z.~Yousaf, and Zarnoor {\em Gen. Relativ. Gravit.}, vol.~51,
  p.~114, 2019.

\bibitem{yousaf2020definition}
Z.~Yousaf, ``Definition of complexity factor for self-gravitating systems in
  palatini $f (\textsc{R})$ gravity,'' {\em Phys Scr.}, vol.~95, p.~075307,
  2020.

\bibitem{herrera2012cylindrically}
L.~Herrera, A.~Di~Prisco, and J.~Ospino, ``Cylindrically symmetric relativistic
  fluids: a study based on structure scalars,'' {\em Gen. Relativ. Gravit.},
  vol.~44, p.~2645, 2012.

\bibitem{sharif2012structure}
M.~Sharif and M.~Z. U.~H. Bhatti, ``Structure scalars for charged cylindrically
  symmetric relativistic fluids,'' {\em Gen. Relativ. Gravit.}, vol.~44,
  p.~2811, 2012.

\bibitem{sharif2015radiating}
M.~Sharif and Z.~Yousaf {\em Astrophys. Space Sci.}, vol.~357, no.~1, p.~49,
  2015.

\bibitem{bhatti2021structure}
M.~Bhatti, Z.~Yousaf, and Z.~Tariq, ``Structure scalars and their evolution for
  massive objects in f (r) gravity,'' {\em Eur. Phys. J. C}, vol.~81, p.~16,
  2021.

\bibitem{yousaf2020evolution}
Z.~Yousaf, M.~Z. Bhatti, and T.~Naseer, ``Evolution of the charged dynamical
  radiating spherical structures,'' {\em Ann. Phys.}, vol.~420, p.~168267,
  2020.

\bibitem{bhatti2020spherical}
M.~Z. Bhatti, Z.~Yousaf, and M.~Nawaz, ``Spherical collapse with heat
  dissipation in f (r, t, r $\mu$ $\nu$ t $\mu$ $\nu$) gravity,'' {\em Int. J.
  Geom. Meth. Mod. Phys.}, vol.~17, p.~2050017, 2020.

\bibitem{yousaf2019non}
Z.~Yousaf, M.~Z. Bhatti, and M.~F. Malik {\em Eur. Phys. J. Plus}, vol.~134,
  p.~470, 2019.

\bibitem{herrera2009dynamics}
L.~Herrera, A.~Di~Prisco, E.~Fuenmayor, and O.~Troconis, ``Dynamics of viscous
  dissipative gravitational collapse: a full causal approach,'' {\em Int. J.
  Mod. Phys. D}, vol.~18, no.~01, pp.~129--145, 2009.

\bibitem{herrera2014shear}
L.~Herrera, A.~Di~Prisco, and J.~Ospino, ``Shear-free axially symmetric
  dissipative fluids,'' {\em Phys. Rev. D .}, vol.~89, no.~12, p.~127502, 2014.

\bibitem{govender2014role}
M.~Govender, K.~Reddy, and S.~Maharaj, ``The role of shear in dissipative
  gravitational collapse,'' {\em Int. J. Mod. Phys. D}, vol.~23, no.~02,
  p.~1450013, 2014.

\bibitem{reddy2015impact}
K.~P. Reddy, M.~Govender, and S.~Maharaj, ``Impact of anisotropic stresses
  during dissipative gravitational collapse,'' {\em Gen. Relativ. Gravit},
  vol.~47, no.~4, p.~35, 2015.

\bibitem{yousaf2020construction}
Z.~Yousaf {\em Phys. Dark Universe}, vol.~28, p.~100509, 2020.

\bibitem{yousaf2020gravastars}
Z.~Yousaf, M.~Z. Bhatti, and H.~Asad {\em Phys. Dark Universe}, vol.~28,
  p.~100527, 2020.

\bibitem{sahoo2017wormholes}
P.~K. Sahoo, P.~H. R.~S. Moraes, and P.~Sahoo {\em arXiv preprint
  arXiv:1709.07774}, 2017.

\bibitem{bambi2016wormholes}
C.~Bambi, A.~Cardenas-Avendano, G.~J. Olmo, and D.~Rubiera-Garcia {\em Phys.
  Rev. D}, vol.~93, p.~064016, 2016.

\bibitem{yadav2020existence}
A.~K. Yadav, L.~K. Sharma, B.~K. Singh, and P.~K. Sahoo {\em New Astr.},
  vol.~78, p.~101382, 2020.

\bibitem{malik2020study}
A.~Malik and M.~F. Shamir {\em New Astr.}, vol.~80, p.~101422, 2020.

\bibitem{yousaf2017stellar}
Z.~Yousaf {\em Eur. Phys. J. Plus}, vol.~132, no.~6, p.~276, 2017.

\bibitem{joshi1992structure}
P.~Joshi and I.~Dwivedi, ``The structure of naked singularity in self-similar
  gravitational collapse,'' {\em Commun. Math. Phys}, vol.~146, no.~2,
  pp.~333--342, 1992.

\bibitem{sussman2009quasilocal}
R.~A. Sussman, ``Quasilocal variables in spherical symmetry: Numerical
  applications to dark matter and dark energy sources,'' {\em Phys. Rev. D .},
  vol.~79, no.~2, p.~025009, 2009.

\bibitem{herrera2011tilted}
L.~Herrera, A.~Di~Prisco, and J.~Ib{\'a}{\~n}ez {\em Phys. Rev. D}, vol.~84,
  no.~6, p.~064036, 2011.

\bibitem{zibin2008scalar}
J.~P. Zibin, ``Scalar perturbations on lemaitre-tolman-bondi spacetimes,'' {\em
  Phys. Rev. D .}, vol.~78, no.~4, p.~043504, 2008.

\bibitem{fernandes2020high}
R.~L. Fernandes, E.~M. Abreu, and M.~B. Ribeiro, ``High-derivatives and massive
  electromagnetic models in the lema{\^\i}tre--tolman--bondi spacetime,'' {\em
  Eur. Phys. J. C .}, vol.~80, no.~3, pp.~1--11, 2020.

\bibitem{herrera2011physical}
L.~Herrera, ``physical causes of energy density inhomogenization and stability
  of energy density homogeneity in relativistic self-gravitaing fluids,'' {\em
  Int. J. Mod. Phys. D}, vol.~20, p.~1689, 2011.

\bibitem{sharif2015stability}
M.~Sharif and Z.~Yousaf, ``Stability of regular energy density in palatini $ f
  (\textsc{R}) $ gravity,'' {\em Eur. Phys. J. C}, vol.~75, p.~1, 2015.

\bibitem{herrera2010lemaitre}
L.~Herrera, A.~Di~Prisco, J.~Ospino, and J.~Carot, ``Lemaitre-tolman-bondi dust
  spacetimes: Symmetry properties and some extensions to the dissipative
  case,'' {\em Phys. Rev. D}, vol.~82, p.~024021, 2010.

\bibitem{yousaf2016causes}
Z.~Yousaf, K.~Bamba, and M.~Bhatti, ``Causes of irregular energy density in $f
  (\textsc{R}, \textsc{T})$ gravity,'' {\em Phys. Rev. D}, vol.~93, p.~124048,
  2016.

\bibitem{herrera2015gravitational}
L.~Herrera, ``Gravitational radiation within its source,'' {\em arXiv preprint
  arXiv:1506.02422}, 2015.

\bibitem{olmo2005post}
G.~J. Olmo, ``Post-newtonian constraints on $ f (\textsc{R})$ cosmologies in
  metric and palatini formalism,'' {\em Phys. Rev. D}, vol.~72, p.~083505,
  2005.

\bibitem{olmo2012stellar}
G.~J. Olmo, H.~Sanchis-Alepuz, and S.~Tripathi, ``Stellar structure equations
  in extended palatini gravity,'' {\em Phys. Rev. D}, vol.~86, p.~104039, 2012.

\bibitem{capozziello2012wormholes}
S.~Capozziello, T.~Harko, T.~S. Koivisto, F.~S. Lobo, and G.~J. Olmo,
  ``Wormholes supported by hybrid metric-palatini gravity,'' {\em Phys. Rev.
  D}, vol.~86, p.~127504, 2012.

\bibitem{olmo2011palatini}
G.~J. Olmo, ``Palatini approach to modified gravity: $f (\textsc{R})$ theories
  and beyond,'' {\em Int. J. Mod. Phys. D}, vol.~20, p.~413, 2011.

\bibitem{amarzguioui2006cosmological}
M.~Amarzguioui, {\O}.~Elgar{\o}y, D.~F. Mota, and T.~Multam{\"a}ki,
  ``Cosmological constraints on $f (\textsc{R})$ gravity theories within the
  palatini approach,'' {\em Astronomy \& Astrophysics}, vol.~454, p.~707, 2006.

\bibitem{allemandi2006conformal}
G.~Allemandi, M.~Capone, S.~Capozziello, and M.~Francaviglia, ``Conformal
  aspects of the palatini approach in extended theories of gravity,'' {\em Gen
  Relativ Gravit.}, vol.~38, p.~33, 2006.

\bibitem{meng2004palatini}
X.~Meng and P.~Wang, ``Palatini formulation of modified gravity with squared
  scalar curvature,'' {\em Gen Relativ Gravit.}, vol.~36, p.~2673, 2004.

\bibitem{misner1964relativistic}
C.~W. Misner and D.~H. Sharp, ``Relativistic equations for adiabatic,
  spherically symmetric gravitational collapse,'' {\em Phys. Rev.}, vol.~136,
  no.~2B, p.~B571, 1964.

\bibitem{PhysRevD.88.064015}
J.~M.~M. Senovilla, ``Junction conditions for $f(r)$ gravity and their
  consequences,'' {\em Phys. Rev. D}, vol.~88, p.~064015, Sep 2013.

\bibitem{sharif2014dynamical}
M.~Sharif and Z.~Yousaf, ``Dynamical analysis of radiating spherical collapse
  in palatini $f (\textsc{R})$ gravity,'' {\em Astrophys. Space Sci}, vol.~354,
  p.~481, 2014.

\bibitem{bel1961inductions}
L.~Bel, ``Inductions {\'e}lectromagn{\'e}tique et gravitationnelle,'' in {\em
  Annales de l'institut Henri Poincar{\'e}}, vol.~17, p.~37, 1961.

\bibitem{herrera2009structure}
L.~Herrera, J.~Ospino, A.~Di~Prisco, E.~Fuenmayor, and O.~Troconis, ``Structure
  and evolution of self-gravitating objects and the orthogonal splitting of the
  riemann tensor,'' {\em Phys. Rev. D}, vol.~79, p.~064025, 2009.

\bibitem{bhatti2021role}
M.~Bhatti, Z.~Yousaf, and Z.~Tariq, ``Role of structure scalars on the
  evolution of compact objects in palatini $f (\textsc{R})$ gravity,'' {\em
  Chinese J. Phys.}, vol.~72, p.~18, 2021.

\bibitem{bhatti2021analysis}
M.~Bhatti, Z.~Yousaf, and Z.~Tariq, ``Analysis of structure scalars in $ f
  (\textsc{R}) $ gravity with an electric charge,'' {\em Phys Scr.}, 2021.

\bibitem{herrera2015physical}
L.~Herrera, A.~Di~Prisco, and J.~Ospino, ``Physical infeasibility of geodesic
  dissipative dust as a source of gravitational radiation,'' {\em Phys. Rev.
  D}, vol.~91, p.~124015, 2015.

\end{thebibliography}
\end{document}